\documentclass[usenatbib]{mn2e}
\usepackage{graphicx,ifthen,url}
\def\ltsima{$\; \buildrel < \over \sim \;$}
\def\simlt{\lower.5ex\hbox{\ltsima}}
\def\gtsima{$\; \buildrel > \over \sim \;$}
\def\simgt{\lower.5ex\hbox{\gtsima}}
\newcommand {\uJy}{$\mu$Jy}
\newcommand {\um}{$\mu$m}

\newcommand {\Mpy}{M$_\odot$\,yr$^{-1}$}
\newcommand {\aj}{AJ}
\newcommand {\aap}{A\&A}
\newcommand {\apj}{ApJ}
\newcommand {\apjs}{ApJS}
\newcommand {\apjl}{ApJL}
\newcommand {\mnras}{MNRAS}

\newcommand {\nat}{Nature}

\def\um     {$\mu$m}
\def\ts     {\thinspace}
\def\kms    {\ifmmode{{\rm \ts km\ts s}^{-1}}\else{\ts km\ts s$^{-1}$}\fi}
\def\msol   {\ifmmode{{\rm M}_{\odot}}\else{M$_{\odot}$}\fi}
\def\lsol   {\ifmmode{{\rm L}_{\odot}}\else{L$_{\odot}$}\fi}
\def\zsol   {\ifmmode{{\rm Z}_{\odot}}\else{Z$_{\odot}$}\fi}
\def\etal   {{\rm et\ts al.}}
\def\ci     {\ifmmode{{\rm C}{\rm \small I}}\else{C\ts {\scriptsize I}}\fi}
\def\hi     {\ifmmode{{\rm H}{\rm \small I}}\else{H\ts {\scriptsize I}}\fi}
\def\hh     {\ifmmode{{\rm H}_2}\else{H$_2$}\fi}
\def\cone {\ifmmode{{\rm C}{\rm \small I}(^3\!P_1\!\to^3\!P_0)}
     \else{C\ts {\scriptsize I}{\small$(^3\!P_1\!\to\,^3\!P_0)$}}\fi}
\def\ctwo {\ifmmode{{\rm C}{\rm \small I}(^3\!P_2\!\to\,^3\!P_1)}
     \else{C\ts {\scriptsize I}{\small$(^3\!P_2\!\to\,^3\!P_1)$}}\fi}
\def\cij {\ifmmode{{\rm C}{\rm \small I}\,(^3P_i\to^3P_j)}\else{C\ts {\scriptsize I}\,{\small$(^3P_i\to^3P_j)$}}\fi}
\def\cii    {\ifmmode{{\rm C}{\rm \small II}}\else{C\ts {\scriptsize II}}\fi}
\def\tex {\ifmmode{{T}_{\rm ex}}\else{$T_{\rm ex}$}\fi}
\def\tmb {\ifmmode{{T}_{\rm mb}}\else{$T_{\rm mb}$}\fi}
\def\tkin {\ifmmode{{T}_{\rm kin}}\else{$T_{\rm kin}$}\fi}
\def\microns {\ifmmode{\mu{\rm m}}\else{$\mu$m}\fi}
\def\nhh   {\ifmmode{n({\rm H}_2)}\else{$n$(H$_2$)}\fi}

\newcommand{\msun}{{\rm\,M$_\odot$}}

\newcommand{\lsun}{{\rm\,L$_\odot$}}

\newcommand{\Ha}{{\rm\,H$\alpha$}}
\newcommand{\Nii}{{\rm\,[N{\sc II}]}}
\newcommand{\Oii}{{\rm\,[O{\sc II}]}}
\newcommand{\Oiii}{{\rm\,[O{\sc III}]}}

\loadboldmathitalic 
\title [250\um\ Galaxies at z$>$1]
{Spectroscopic Characterisation of 250\um-Selected Hyper-Luminous Star Forming Galaxies}
\author[C.~M. Casey et al.]
{
C.~M. Casey$^{1,2}$\thanks{Hubble Fellow; cmcasey@ifa.hawaii.edu}, S.~C. Chapman$^1$, Ian Smail$^3$, S. Alaghband-Zadeh$^{1}$, 
\newauthor M.~S. Bothwell$^{1}$, A.~M. Swinbank$^3$\\
$^1$ Institute of Astronomy, Madingley Road, Cambridge, CB3 0HA \\
$^2$ Institute for Astronomy, University of Hawai'i, 2680 Woodlawn Dr, Honolulu, HI 96822, USA \\
$^3$ Institute for Computational Cosmology, Durham University, South Road, Durham DH1 3LE \\
}
\date{Submitted 12 July 2010, Revised with minor corrections from referee, 10 September 2010.  Version date \today}
\pagerange{\pageref{firstpage}--\pageref{lastpage}} \pubyear{2009}

\begin{document} 
\maketitle 
\label{firstpage}

\begin{abstract}
  We present near-infrared spectroscopic observations from VLT {\sc
    ISAAC} of thirteen 250\um-luminous galaxies in the CDF-S, seven of
    which have confirmed redshifts which average to $\langle z
    \rangle$\,=\,2.0$\pm$0.4.  Another two sources of the 13 have
    tentative $z\,>$\,1 identifications.  Eight of the nine redshifts
    were identified with H$\alpha$ detection in H- and K-bands, three
    of which are confirmed redshifts from previous spectroscopic
    surveys.  We use their near-IR spectra to measure \Ha\ line widths
    and luminosities, which average to 415$\pm$20\kms\ and
    3$\times$10$^{35}$\,W (implying SFR$_{H\alpha}\,\sim$\,200\,\Mpy),
    both similar to the \Ha\ properties of SMGs.  Just like SMGs,
    250\um-luminous galaxies have large \Ha\ to far-infrared (FIR)
    extinction factors such that the \Ha\ SFRs underestimate the FIR
    SFRs by $\sim$8-80\,times.  Far-infrared photometric points from
    observed 24\um\ through 870\um\ are used to constrain the spectral
    energy distributions (SEDs) even though uncertainty caused by FIR
    confusion in the BLAST bands is significant.  The population has a
    mean dust temperature of T$_d$\,=\,52$\pm$6\,K, emissivity
    $\beta$\,=\,1.73$\pm$0.13, and FIR luminosity $L_{\rm
    FIR}$\,=\,3$\times$10$^{13}$\,\lsun.
%Six are also 870\um\
%    detected, thus classified as submillimetre galaxies (SMGs), but
%    the rest is submm-faint ($<$5\,mJy at 870\um), indicating that
%    this population is not entirely overlapping with the SMG
%    population.  
Although selection at 250\um\ allows for the detection of much hotter
    dust dominated HyLIRGs than SMG selection (at 850\um), we do not
    find any \simgt60\,K `hot-dust' HyLIRGs.  
%Since the temperature
%    distribution in local ULIRGs would suggest detection of at least
%    one \simgt60\,K high-$z$ source (with $\sim$50\%\ likelihood),
%    this is possible evidence that extreme starbursts at high-$z$ are
%    more extended than local systems, thus have colder, more diffuse
%    dust.  
We have shown that near-infrared spectroscopy combined with good
    photometric redshifts is an efficient way to spectroscopically
    identify and characterise these rare, extreme systems, hundreds of
    which are being discovered by the newest generation of IR
    observatories including the {\it Herschel Space Observatory}.

% which has significant implications for the cosmic star
%  formation rate density at z$\sim$2 and the nature of extreme starbursts
\end{abstract}

\section{Introduction}\label{s:introduction}

Submillimetre Galaxies (SMGs) contribute significantly to the rapid
buildup of stellar mass in the Universe at $z\,\sim\,$2.  However,
their selection at 850\um\ is inherently biased towards colder-dust
sources \citep{eales00a,blain04a}.  Recent work
\citep[e.g.][]{chapman04a,casey09b} has demonstrated that
850\um-faint, high-redshift Ultraluminous Infrared Galaxies (ULIRGs)
exist and may contribute significantly to the cosmic star formation
rate density at its peak.  \citet{casey09b} describe a population of
70\um\ luminous galaxies at $z\sim$1.5 whose infrared luminosities
exceed $\sim$10$^{12}$\,\lsun\ but are 850\um-faint due to hotter
characteristic dust temperatures.  Sparse infrared data, particularly
in the 50-500\um\ wavelength range, along with poor volume density
constraints have limited the interpretation of these submm-faint
ULIRGs.  Similar studies of other infrared-luminous galaxy
populations, selected at 24\um, 350\um\ or 1.2mm for example, present
even more evidence for diverse populations of luminous, dusty
starbursts at z\simgt1 which do not necessarily intersect
\citep[see][]{dey08a,bussmann09a,younger09a}.

The arrival of new infrared (IR) instruments$-$including BLAST
\citep[the Balloon-borne Large-Aperture Submillimeter
Telescope][]{pascale08a}, SCUBA2, LABOCA, and the {\it Herschel Space
Observatory}$-$has opened up more extensive studies of these distant
starbursts.  BLAST's deep mapping of the Extended {\it Chandra} Deep
Field South (ECDF-S) at 250\um, 350\um, and 500\um\ has, for the first
time, led to z$\sim$2 ULIRG selection near the peak of their SED.
\citet{ivison10a} and \citet{dunlop10a} describe the selection of
these 250\um\ sources, along with their radio and 24\um\ counterparts,
in detail and match sources to photometric redshifts derived from the
extensive ECDF-S multi-wavelength data.  We also make use of longer
wavelength constraints from the LABOCA 870\um\ survey of the ECDF-S
\citep{weiss09b}.  While most low redshift (z\simlt0.8) 250\um\
sources have spectroscopic identifications, none of the suspected
high-redshift sources had spectroscopic redshifts.

This paper presents new VLT ISAAC spectroscopic observations of
thirteen BLAST 250\um\ sources with $z_{\rm phot}$\,$>$\,1.  With
spectroscopic redshifts, we constrain the FIR dust SED (implying that
they are HyLIRGs with $L_{\rm FIR}$\,\simgt\,10$^{13}$\,\lsun),
measure dust temperatures, blackbody emissivity, FIR luminosities,
\Ha\ luminosities and AGN/metal lines, and constrain the FIR/radio
correlation for these high-$z$ galaxies.  Throughout we use a
$\Lambda_{\rm CDM}$ cosmology \citep{hinshaw09a} with
$H_{0}$=\,71\,\kms\,Mpc$^{-1}$ and $\Omega_{0}\,=\,$0.27.

%{\it Extra bit: delete or expand} Some studies
%(e.g. Sajina et al. 200?) suggest that the cold dust temperatures of
%SMGs are an inherent physical property and not due exclusively to
%selection bias, although testing this hypothesis has been difficult
%without sampling the full ULIRG spectral energy distributions (SEDs)
%from $\sim$50-1000\um.

\begin{figure}
\centering
\includegraphics[width=0.99\columnwidth]{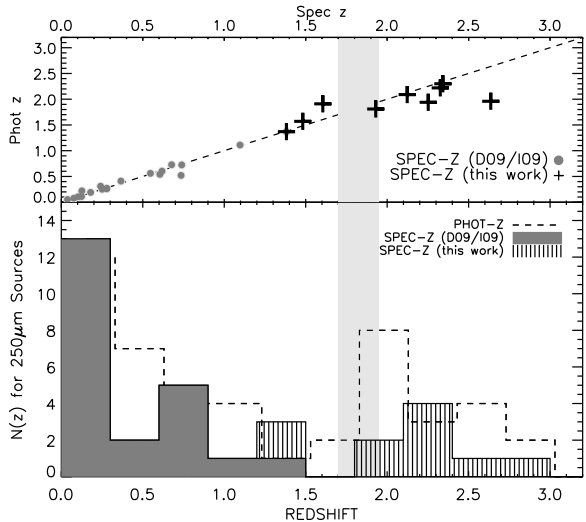}
\caption{ The redshift distribution of BLAST 250\um\ sources,
  including the low-redshift spectroscopically identified sources in
  D10 and I10 ({\it gray points/histogram}).  Our spectroscopic sample
  has a mean redshift 2.0$\pm$0.4 ({\it black crosses, lined
  histogram}), which fills in much of the high-redshift subsample of
  250\um-bright objects. The distribution in photometric redshifts is
  illustrated as the dashed line histogram and are good to $\sim$12\%.
  The light gray vertical stripe blocks out the redshift regime where
  near-IR spectroscopy cannot confirm redshifts since \Ha\ falls
  between H and K bands.}
\label{fig:zdist}
\end{figure}

\section{Observations \& Results}\label{s:observations}

Longslit spectroscopic observations of 250\um\ sources were obtained
in November 2009 on the Very Large Telescope (VLT) Infrared
Spectrometer And Array Camera (ISAAC) under excellent seeing
conditions (0.3-0.7\arcsec\ in K-band).  Spectroscopic candidates were
chosen from the \citet{ivison10a} and \citet{dunlop10a} (hereafter I10
and D10) ECDF-S BLAST Deep map samples (RMS sensitivity
$\sigma_{250}$\,=\,11\,mJy) with photometric redshifts above $z=$1 or
undefined photometric redshifts (the latter caused by a lack of high
quality photometry).  I10 selected sources at $>$5$\sigma$, having
folded in the confusion noise ($\sim$21\,mJy), resulting in flux
densities $S_{250}\,>\,$59\,mJy.  D10 selected sources at $>$3$\sigma$
without accounting for the confusion noise, so their source list has
$S_{250}\,>\,$33\,mJy.  All had reliable 24\um\ and/or radio
counterparts, which were then matched to K-band sources (with offsets
$<$1\arcsec\ to the radio/24\um\ centroid) in archival MUSYC data
\citep{gawiser06a} for VLT spectroscopic targeting.  The resulting
candidate object list contained 20 sources.

We refer the reader to D10 and I10 for the analysis of the BLAST
250\um\ CDF-S map and source selection, as well as some source
properties derived from ancillary data.  The two papers present
different detection thresholds (which are discussed in more detail as
they relate to source density estimations in section
\ref{sec:sourcedensity}) and also use different counterpart
identification methods to identify sources for photometric redshift
fitting.  D10 uses 24\um\ counterpart matching, while I10 uses
1.4\,GHz radio matching.  In general radio matching is much more
reliable assuming the FIR/radio correlation holds \citep{helou85a} as
there are far fewer potential counterparts and good reason to suspect
a FIR-bright source is also radio-bright.  Identification at 24\um\ is
a less reliable alternative due to the large IR beamsize and density
of sources.  We caution the reader that our K-band identifications are
nearest neighbours to the radio and 24\um\ counterpart astrometry of
D10 and I10 and that there is minor potential for misidentified
counterparts.  The offsets between 250\um\ peaks and K-band sources
(which are effectively equivalent to the radio/24\um\ positions) range
from 1-16\arcsec, averaging to about 7\arcsec, which is well within
the beamsize of 250\um\ observations, however it is not possible to
know if the counterparts have been identified correctly without
high-resolution FIR observations \citep[e.g.][]{younger10a}.

We observed 13 of the 20 $z_{\rm phot}>1$ 250\um\ sources searching
for \Ha\ or \Oiii\ in J, H, and K bands.  The band of observations was
primarily chosen based on the galaxies' photometric redshifts (where
galaxies with $z_{phot}>$2 were observed in K-band and at
$z_{phot}<$1.8 in H-band, and in J-band for the intermediate region).
Galaxies were observed individually under varying seeing conditions
which ranged from 0.3-0.7\arcsec\ seeing in K-band.  Since all
galaxies here are assumed to be unresolved, we varied the slit width
according to seeing conditions, minimising it when possible to reduce
sky line contamination.  Galaxies were centred on the 2$'$ long slit
and observed in ABBA nodding mode with a 15\arcsec\ nod.  On occasion,
two candidates were within a 2$'$ separation and placed on the same
slit, with the maximum possible nod distance, which was sometimes
5-10\arcsec.  Data reduction was completed with ESO software combined
with IRAF and our own IDL-based routines to obtain 1D and 2D
wavelength calibrated spectra.

Only 13 of the sources were
observed due to telescope time constraints. Nine of the 13 have
spectroscopic redshifts, seven of which are secure. Three of these
seven sources (J033246, J033152, and J033243) were already identified
in previous spectroscopic surveys using the Very Large Telescope/FOcal
Reducer and low dispersion Spectrograph 2 and the Gemini Near-Infrared
Spectrograph \citep{vanzella08a,kriek08a} at $z\,=\,$1.382, 2.336 and
2.122 respectively.  Our measured redshifts confirm these
observations.  Six of the seven secure redshifts and the two tentative
redshifts were measured from H$\alpha$ detection (at $>$4$\sigma$ \Ha\
signal-to-noise).  One of the eight \Ha\ redshifts (J033129) would
nominally be tentative, but it was spectroscopically identified in the
rest-UV independently; the ninth secure redshift which is not based on
\Ha\ (J033151) has absorption features in the K-band at $\sim$1.599
which agree with a rest-UV redshift of 1.605 obtained independently
(Swinbank, private communication).  The lines were identified as
H$\alpha$ using a combination of photometric redshift consistency and
a lack of other line features (which would instead identify the line
as \Oii\ or \Oiii, in the case of \Nii\ or SII detection).  The four
sources which were not identifiable in emission either have very weak
emission features or lie at redshifts in the range
1.7\,$<$\,$z$\,$<$\,2.0; \Ha\ at these redshifts falls between H- and
K- bands and is thus not detectable with near-IR spectroscopy.

\begin{figure*}
\centering
\includegraphics[width=0.9\columnwidth]{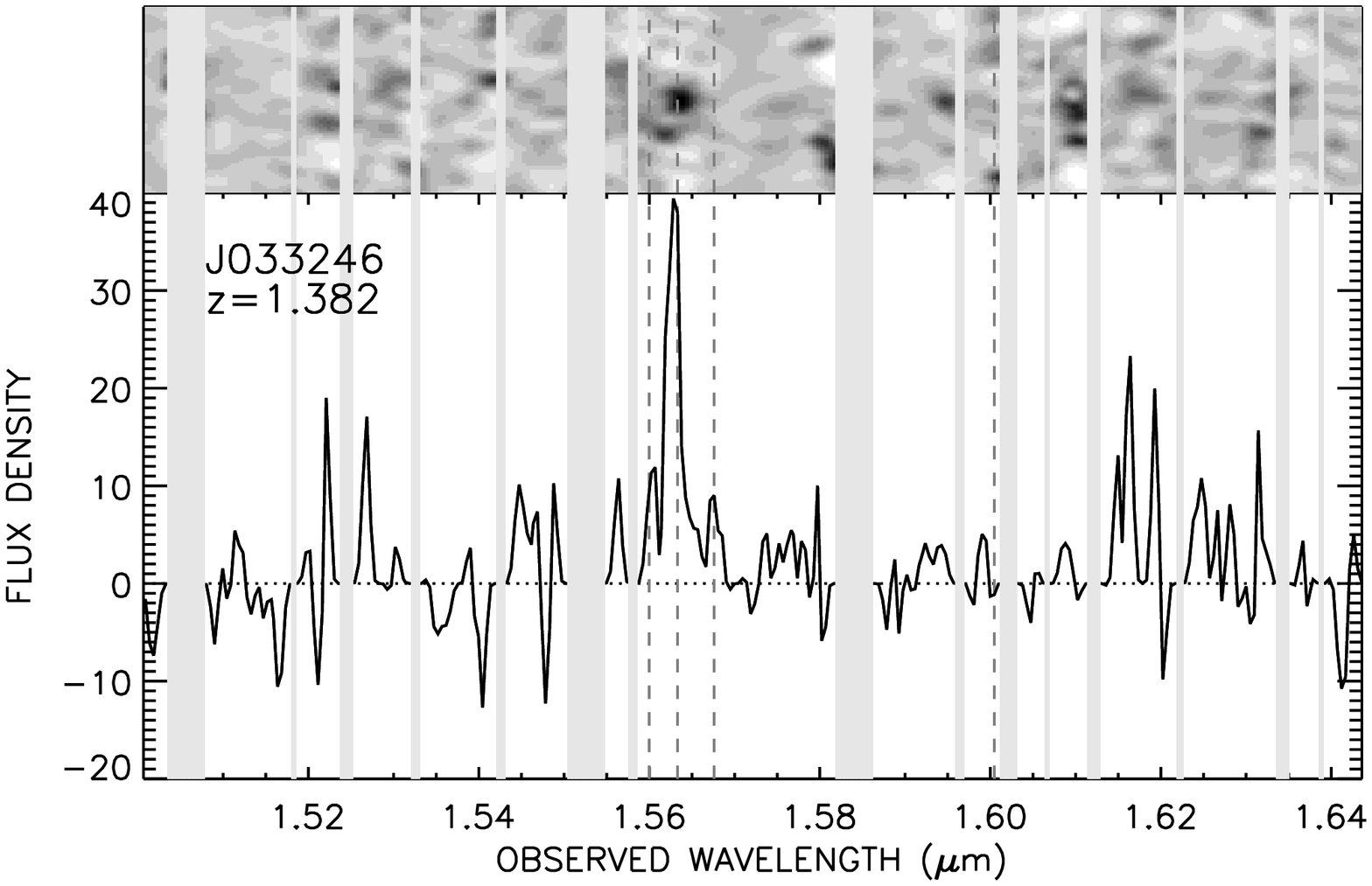}
\includegraphics[width=0.9\columnwidth]{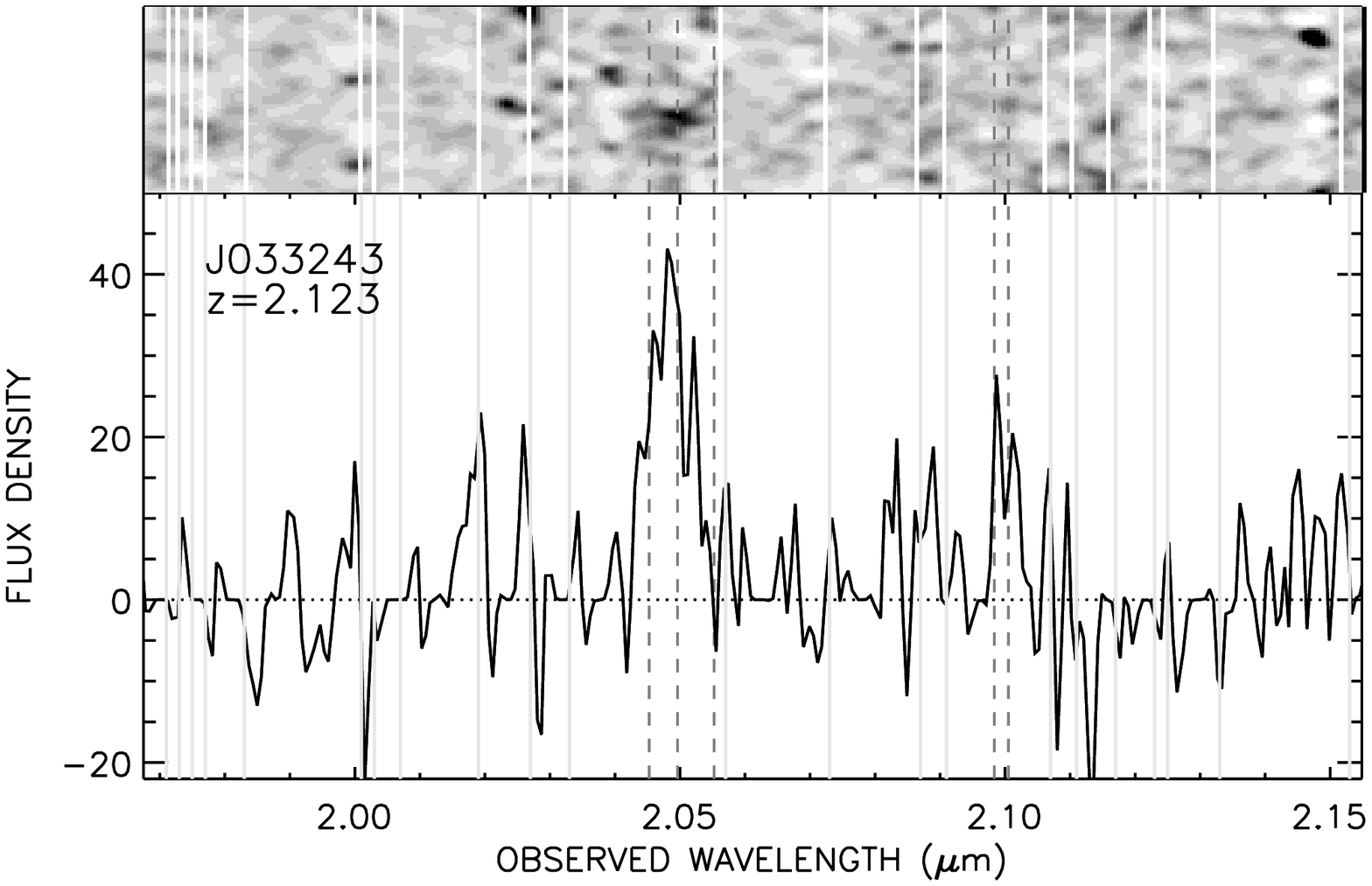}\\
\includegraphics[width=0.9\columnwidth]{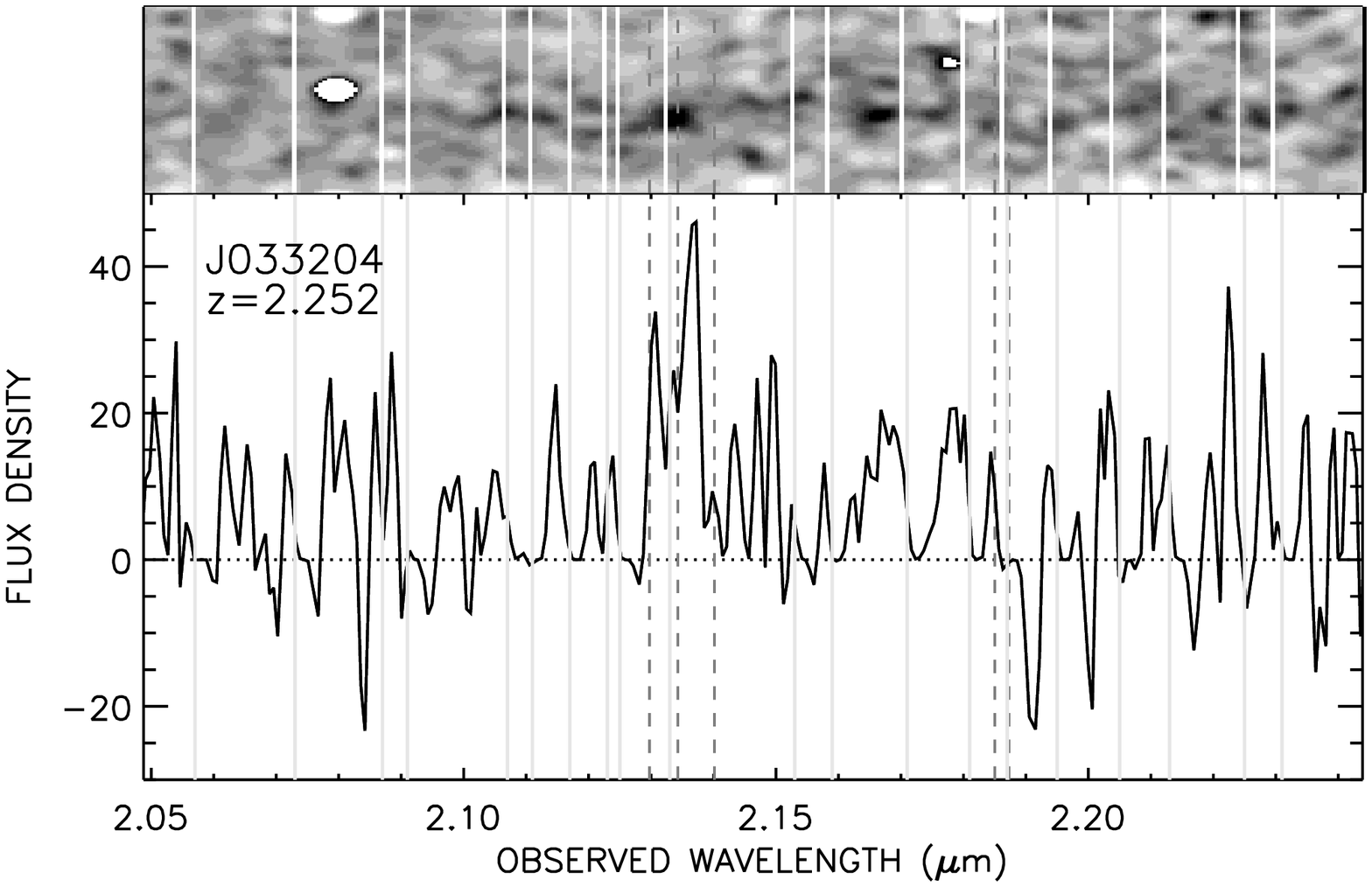}
\includegraphics[width=0.9\columnwidth]{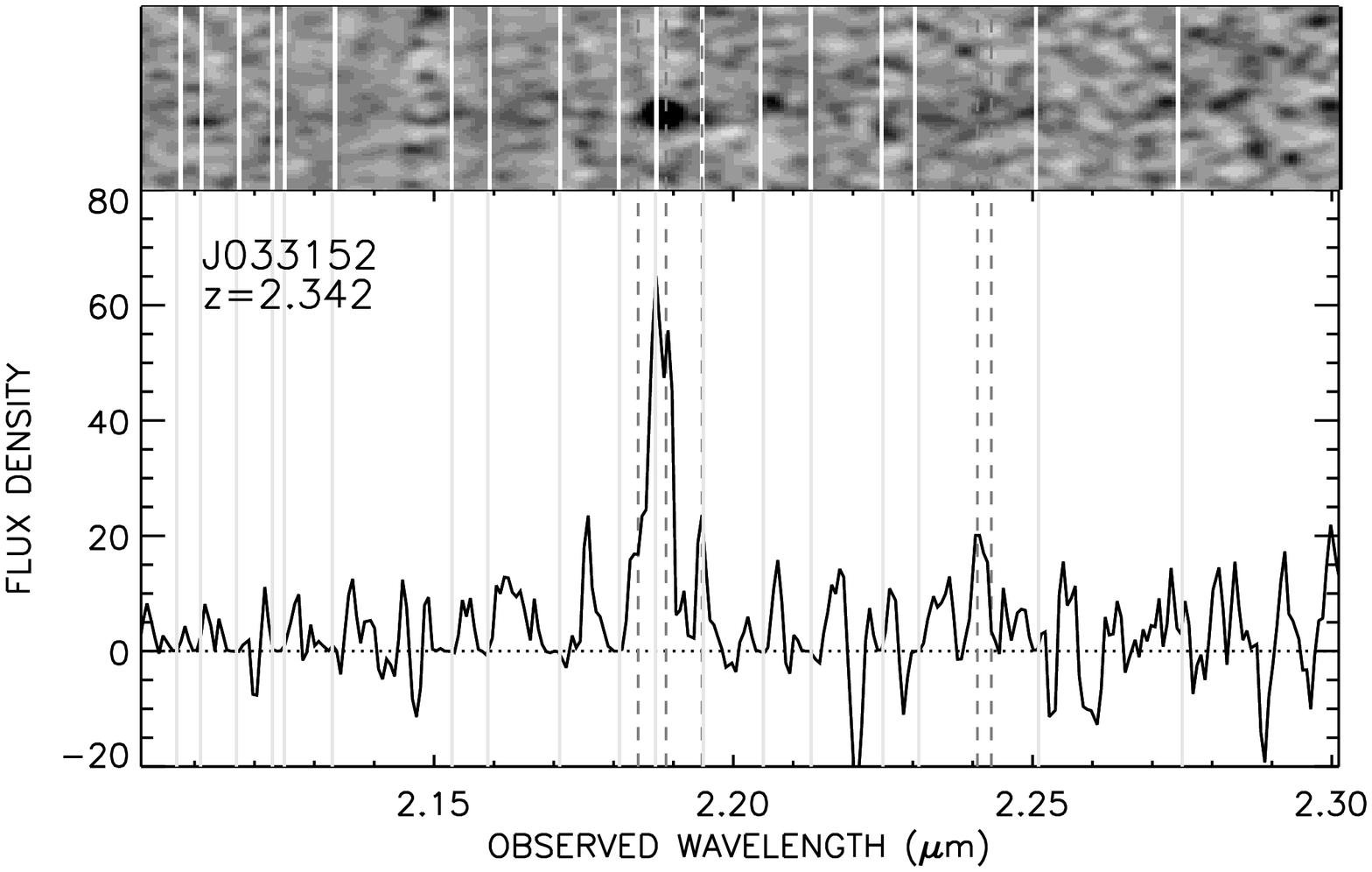}\\
\includegraphics[width=0.9\columnwidth]{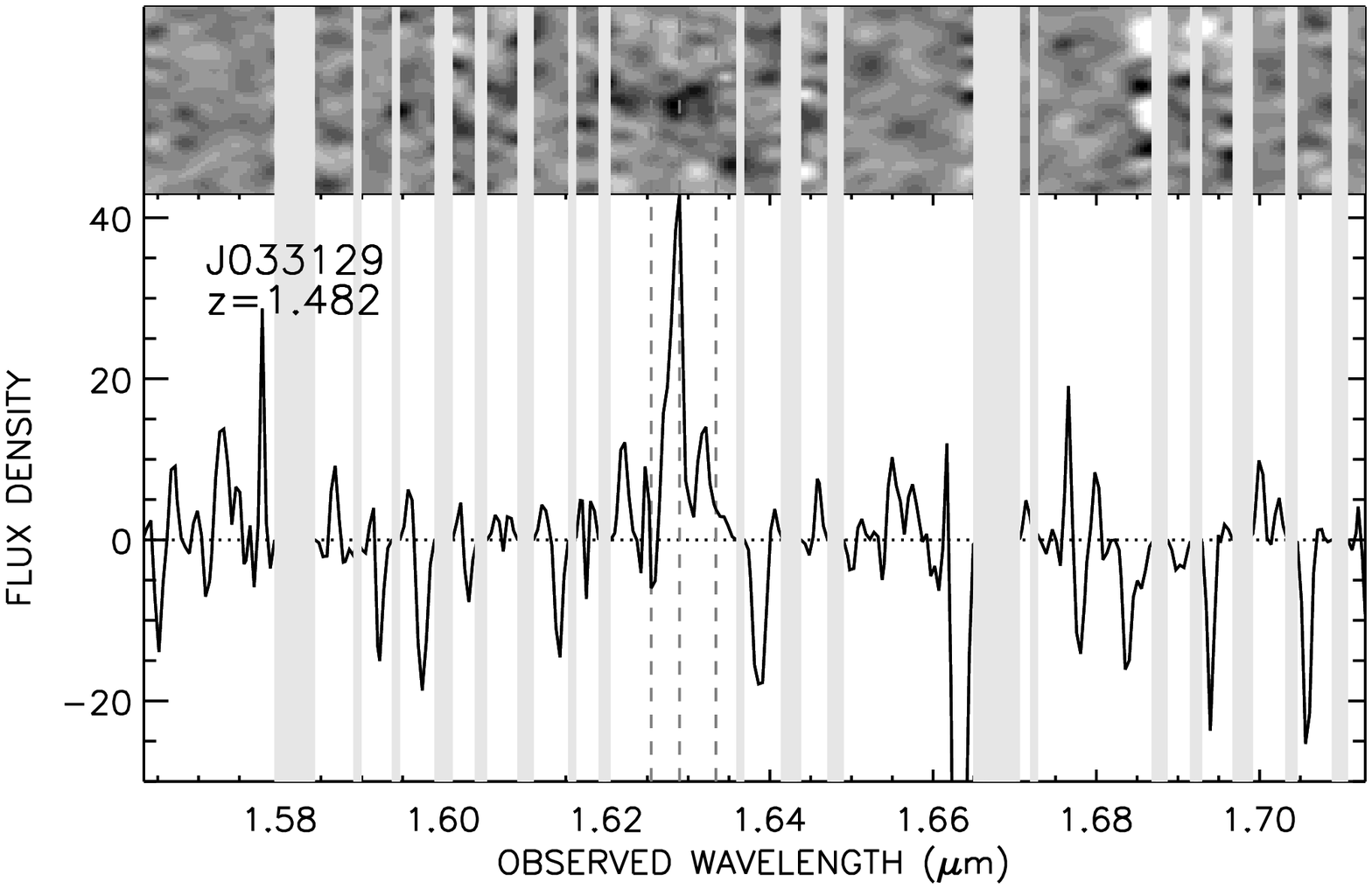}
\includegraphics[width=0.9\columnwidth]{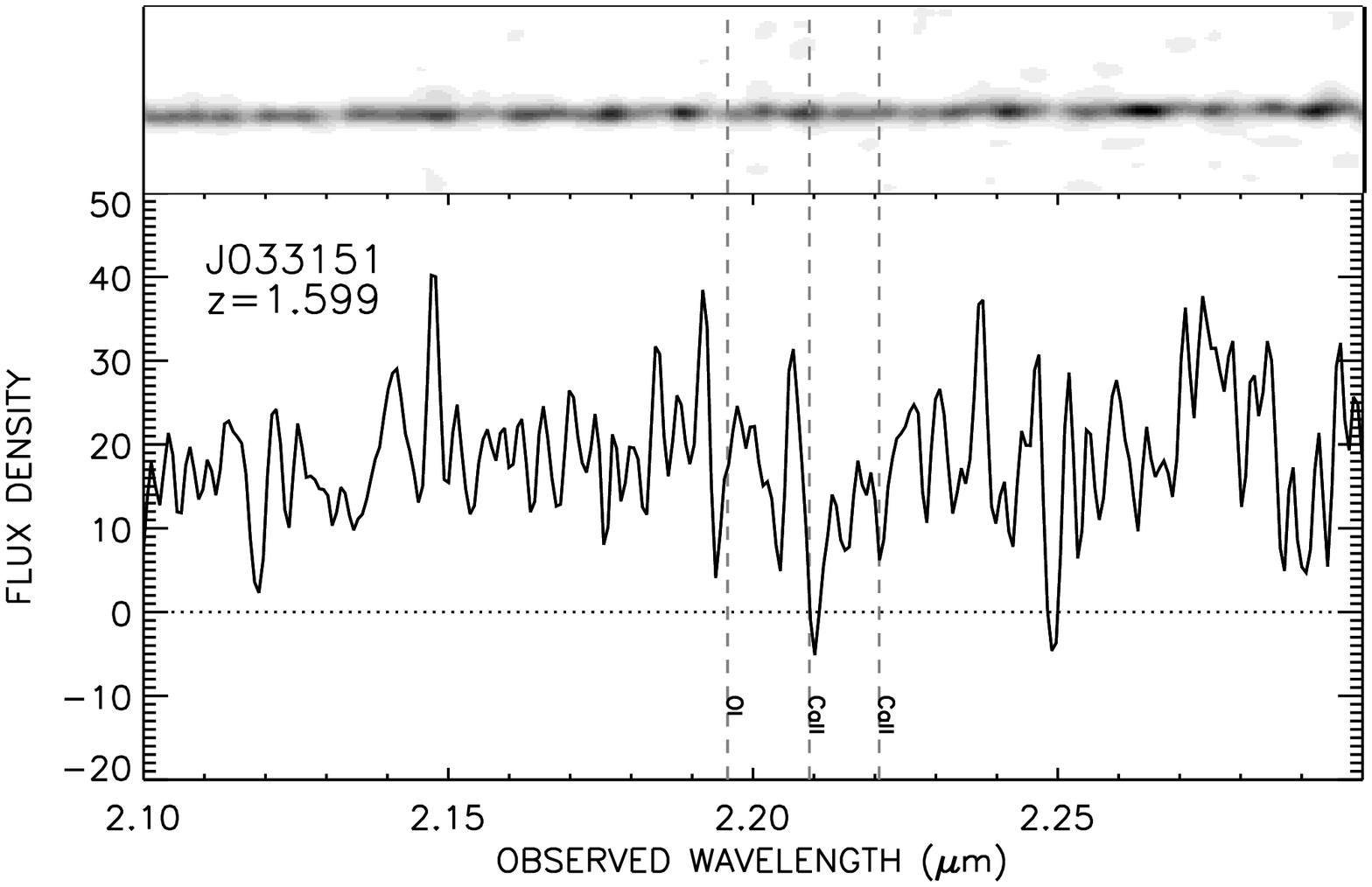}\\
\includegraphics[width=0.9\columnwidth]{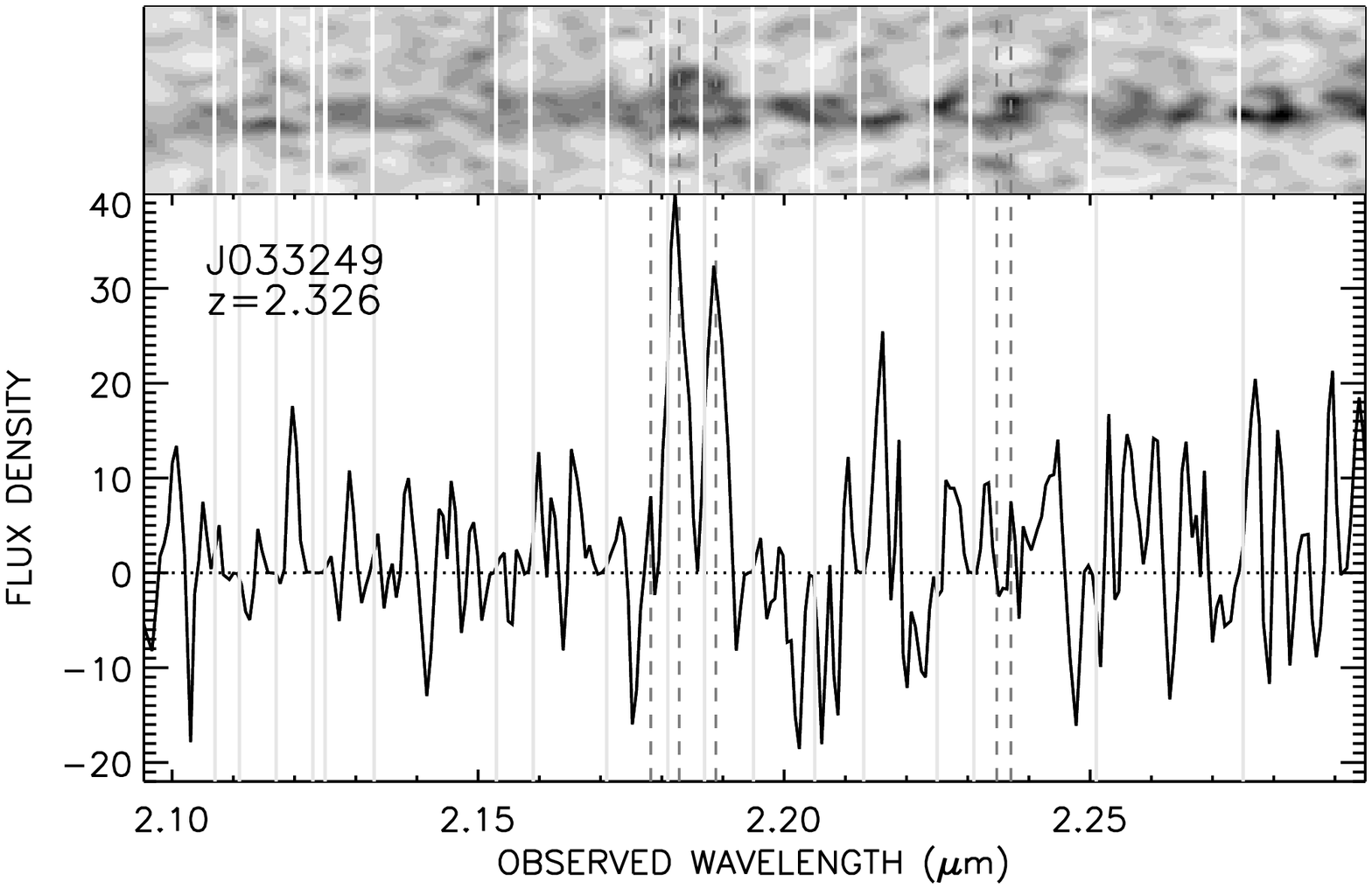}
\includegraphics[width=0.9\columnwidth]{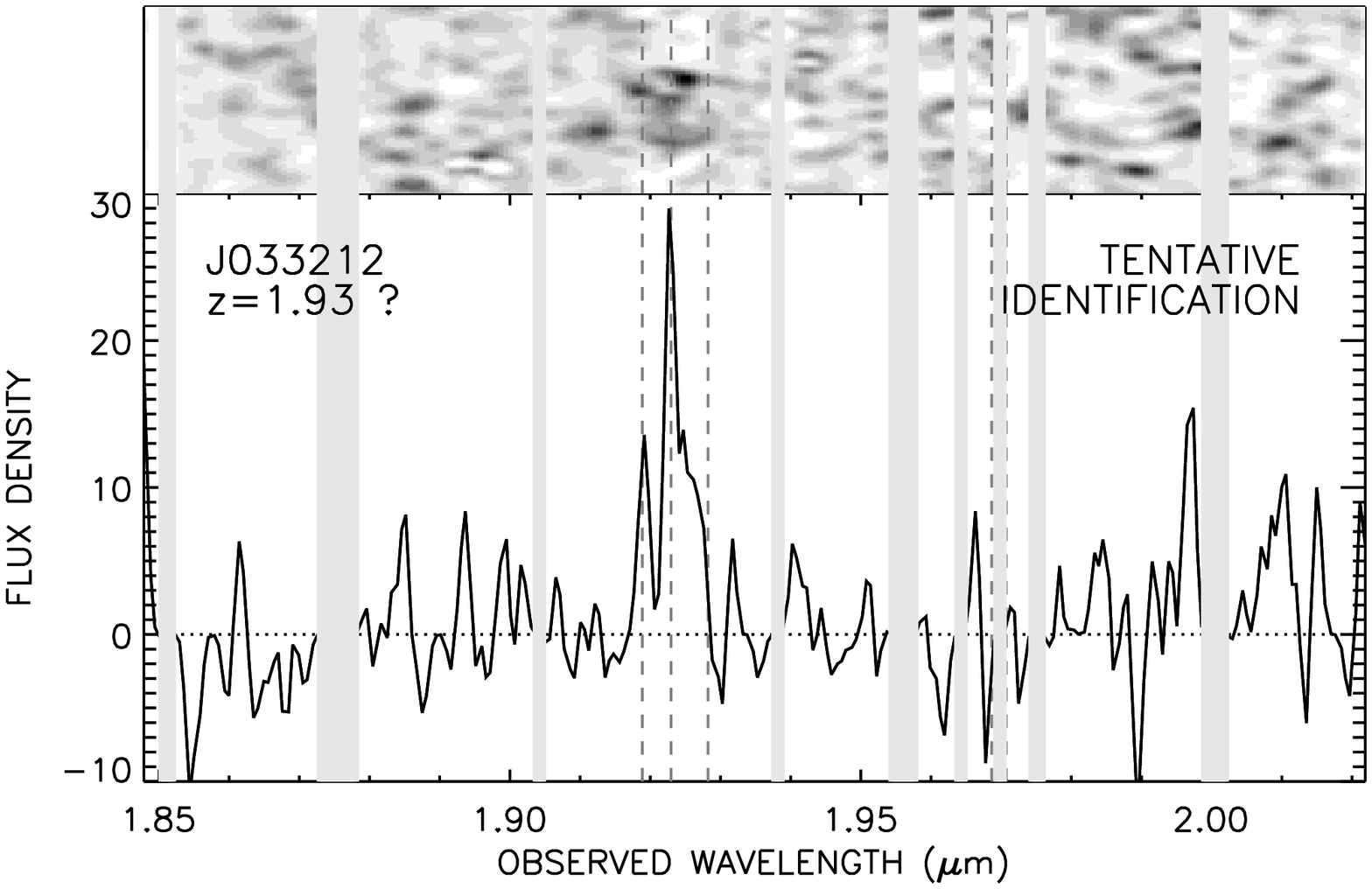}\\
\includegraphics[width=0.9\columnwidth]{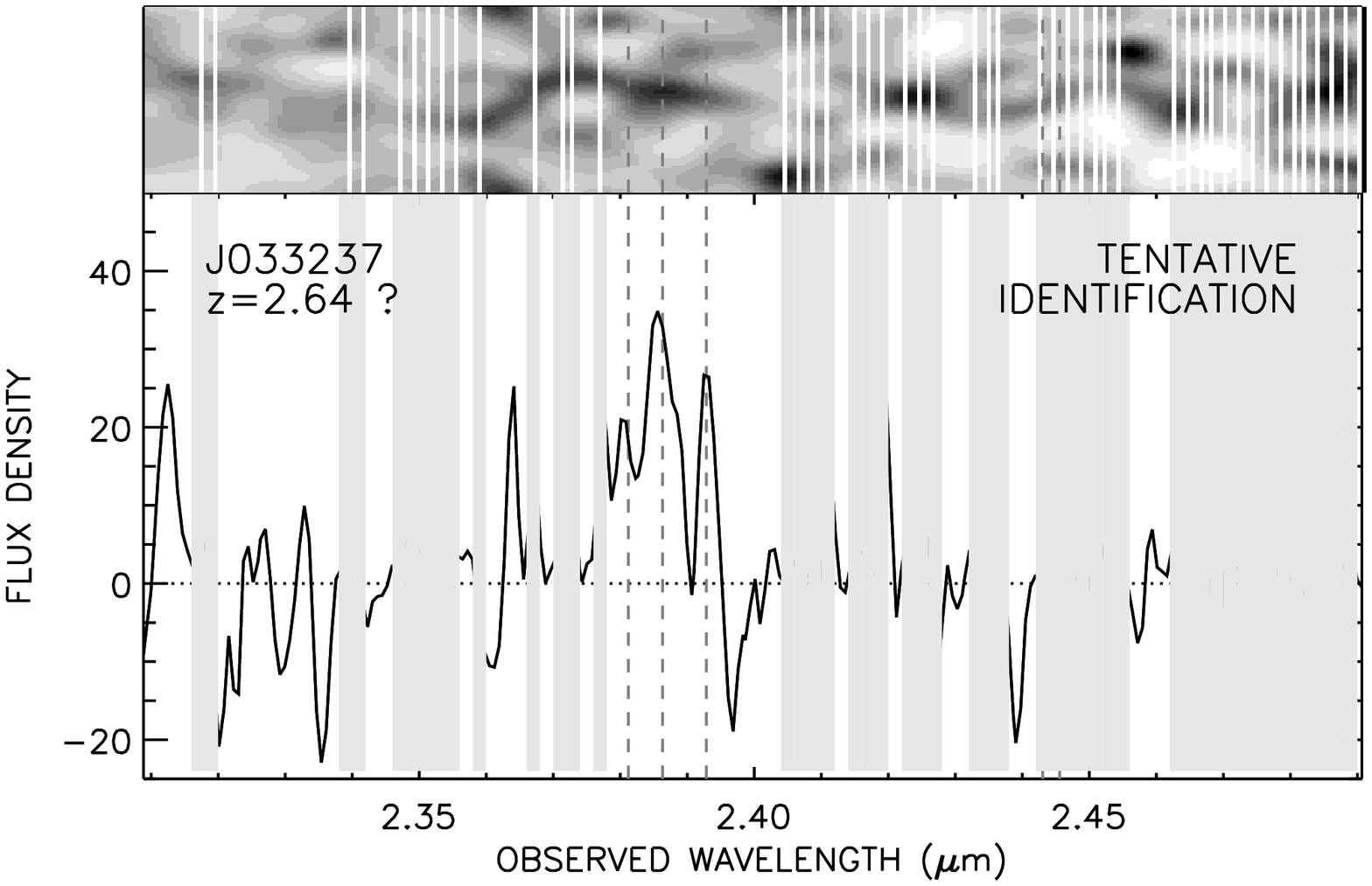}
\includegraphics[width=0.99\columnwidth]{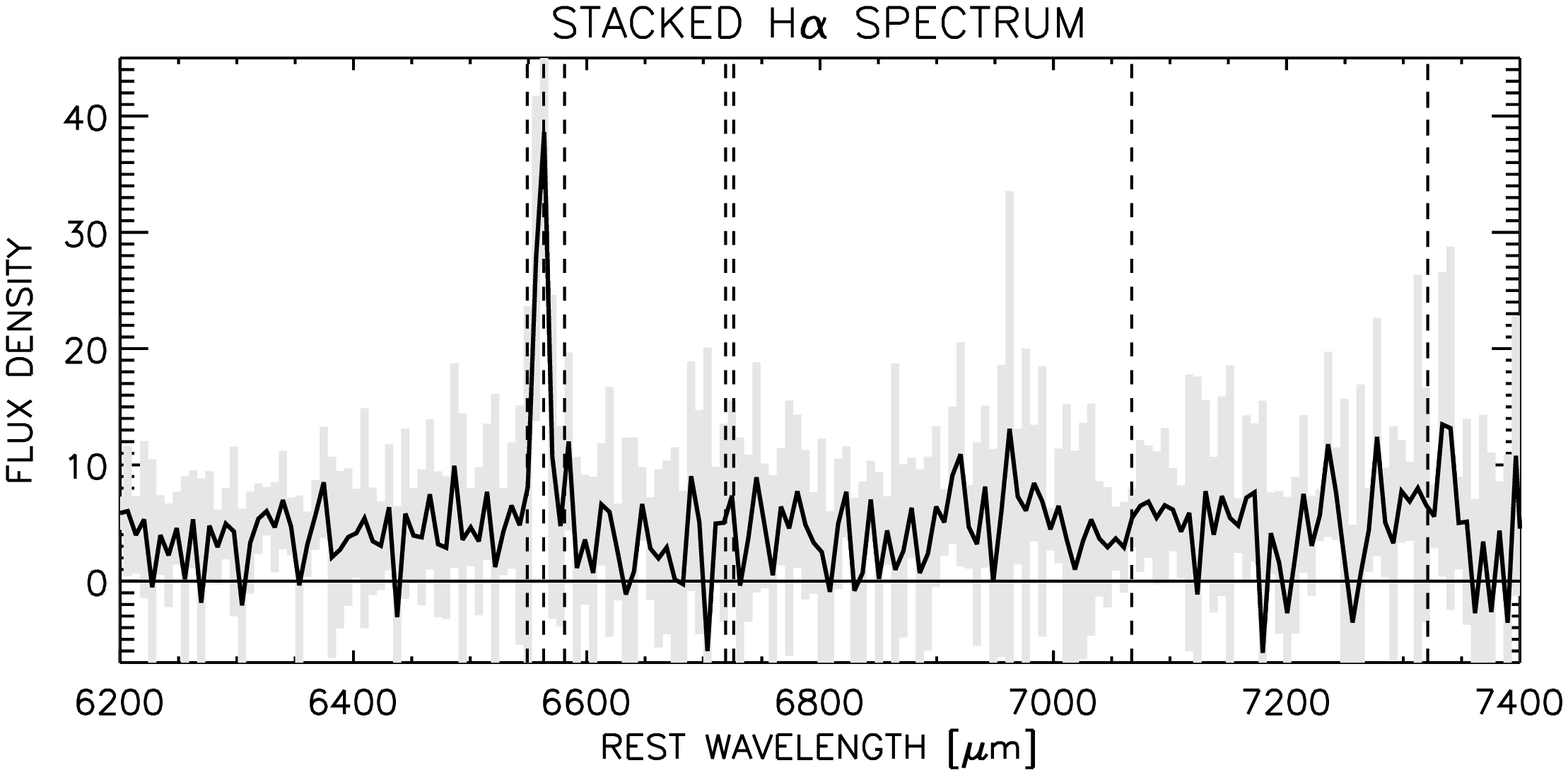}
\caption{ The VLT ISAAC spectroscopy of nine 250\um\ BLAST sources
  around their \Ha\ emission (0.63-0.69\um\ rest wavelength), except
  in the case of J033151 when no \Ha\ observations were obtained.  The
  first seven sources have secure redshifts, the next two are
  tentative, and the last is a stacked \Ha\ spectrum for the six
  secure \Ha\ sources.  Observations were taken in H or K band, and
  both 2d and 1d projections are shown for clarity; scaling is
  optimised for viewing spectral features which are marked by vertical
  dashed lines (e.g. \Ha, NII). The 2d spectra show an angular scale
  of 12\arcsec\ top to bottom and the 1d spectra are extracted within
  0.6\arcsec. Sky line emission features are blocked out (solid gray
  vertical lines).  Higher-redshift sources are smoothed more (e.g. as
  in J033237) since the observed wavelengths (of rest-frame \Ha) are
  higher.  The stacked spectrum for the six secure redshift
  \Ha\ spectra is shown in the lower right.}
\label{fig:isaacspectra}
\end{figure*}

The nine galaxies have a mean redshift $z\,=\,$2.0$\pm$0.4, and their
redshift distribution, with respect to other 250\um\ sources, is shown
in Figure~\ref{fig:zdist}.  We determine that their photometric
redshifts (derived in D10 and in Rafferty \etal, in prep, for I10
sources) are good to $dz$/($1+z$)\,\simlt\,12\%, and we emphasise that
this relatively small error implies that near-IR spectroscopic
followup for ULIRGs with good photometric redshifts is efficient.
Table~\ref{tab:observations} summarises the galaxies properties and
Figure~\ref{fig:isaacspectra} shows their ISAAC spectra for those
which were spectroscopically identified.  The galaxies' names are
derived from their positions in the K band.

The spectra in Figure~\ref{fig:isaacspectra} are framed around the
\Ha\ emission for every source except J033151.  Regions where emission
lines in the sky's infrared spectrum\footnote{See the Gemini
  Observatories IR Background Spectra page
  (http://www.gemini.edu/?q=node/10787) for example sky spectra.} are
significant (with flux densities in excess of
$\sim$5$\gamma$\,s$^{-1}$\,nm$^{-1}$\,arcsec$^{-2}$\,m$^{-2}$ where
$\gamma$ represents photons) are masked out in both 1D and 2D spectral
renditions.  The width of these sky lines varied according to the slit
width of each observation, which varied from 0.3-0.8\arcsec.

\begin{table*}
\begin{center}
\caption{Multi-wavelength properties of BLAST 250\,\um\ Galaxies}
\label{tab:observations}
\begin{tabular}{c@{ }c@{ }c@{ }c@{ }c@{ }c@{ }c@{ }c@{ }c@{ }c@{ }c@{ }c@{  }c@{ }c@{ }c@{ }c}
\hline\hline
NAME$^{a}$ & ID$^{b}$ & $z_{\rm spec}$ & $z_{\rm phot}$ & $S_{24}$ & $S_{250}$ & $S_{350}$$^{d}$ & $S_{500}$$^{d}$ & $S_{1.4}$ & $S_{870}$$^{d}$ & {\sc class}$^{e}$ & L$_{\rm FIR}$ & T$_{\rm dust}$ & $\beta$ & $q_{IR}$$^{f}$ \\
  & & & & (\uJy) & (mJy) & (mJy) & (mJy) & (\uJy) & (mJy) & & (10$^{13}$\lsun) & (K) & & \\
\hline
{\bf \footnotesize DETECTIONS} & & & & & & & & & & & & & & \\
%158
%J033221.624-275623.49 & 2.277 & 1.85 & 510 & 54.9$\pm$10.9 & 28.5$\pm$8.4 & 31.2$\pm$6.0 & 38   & 5.5$\pm$1.3  & (4.1$^{+5.7}_{-3.4}$) & 43.4$\pm$2.6 & =2.0 & 3.31 \\
%iv5  
J033129.874-275722.40 & {\scriptsize J033129} & 1.482 & 1.57 & 270 & 91.6$\pm$11.0 & 54.4$\pm$8.7 & 46.3$\pm$6.2 & 144$\pm$16  & 5.0$\pm$1.5 & SMG & (1.2$^{+0.7}_{-0.4}$) & 45.9$\pm$3.9 & 1.2$\pm$0.3 & 2.7$\pm$0.5 \\
%iv8   
J033151.088-274436.91 & {\scriptsize J033151} & 1.599 & 1.91 & 520 & 74.0$\pm$10.8 & 63.8$\pm$8.5 & 37.4$\pm$5.9 & 96$\pm$13   & 4.6$\pm$1.4 & SMG & (2.2$^{+1.4}_{-0.9}$) & 47.1$\pm$3.0 & 1.6$\pm$0.3 & 3.0$\pm$0.8 \\
%iv3 
J033152.090-273926.32 & {\scriptsize J033152} & 2.342$^{c}$ & 2.30 & 200 & 78.3$\pm$11.0 & 64.3$\pm$8.6 & 53.1$\pm$6.0 & 965$\pm$16  & 2.4$\pm$1.4 & SFRG & (8.1$^{+6.0}_{-3.4}$) & 44.5$\pm$3.0 & 2.6$\pm$0.3 & 2.2$\pm$0.5 \\
%66   
J033204.849-274647.27 & 66 & 2.252 & 1.94 & 540 & 64.3$\pm$10.9 & 62.0$\pm$8.4 & 22.4$\pm$6.0 & 126$\pm$12  & 7.9$\pm$1.4 & SMG  & (4.0$^{+1.0}_{-0.8}$) & 56.7$\pm$5.2 & 1.3$\pm$0.4 & 2.8$\pm$0.6 \\
%318  
J033243.209-275514.38 & 318 & 2.123$^{c}$ & 2.09 & 510 & 30.1$\pm$10.9 & 32.4$\pm$8.5 & 17.8$\pm$6.0 & 92$\pm$10   & 5.7$\pm$1.4 & SMG  & (1.4$^{+2.5}_{-0.9}$) & 56.9$\pm$8.6 & 0.8$\pm$0.7 & 1.6$\pm$0.4 \\
%1293 
J033246.329-275327.01 & 1293 & 1.382$^{c}$ & 1.37 & 200 & 28.1$\pm$ 10.9& 25.3$\pm$8.6 & 14.7$\pm$5.9 & 91$\pm$7  & -1.3$\pm$1.4 & SFRG & (0.4$^{+0.1}_{-0.1}$) & 53.0$\pm$12.9 & 0.8$\pm$0.7 & 2.5$\pm$0.7 \\
%iv2  
J033249.352-275845.07 & {\scriptsize J033249} & 2.326 & 2.22 & 320 & 101.2$\pm$10.9 & 66.4$\pm$8.6 & 22.6$\pm$6.0 & 216$\pm$16 & 2.5$\pm$1.3 & SFRG & (8.1$^{+3.9}_{-2.6}$) & 56.7$\pm$4.5 & 2.1$\pm$0.3 & 2.8$\pm$0.3 \\
%%TENTATIVE DETECTIONS
{\bf \footnotesize TENTATIVE} & & & & & & & & & & & & \\
%193  
J033212.866-274640.89 & 193 & {\it 1.93} & 1.81 & 40  & 46.0$\pm$10.9 & 33.2$\pm$8.5 & 8.6$\pm$6.0 & $<$40 & -0.5$\pm$1.4 & SFRG & (2.2$^{+4.3}_{-1.4}$) & 42.7$\pm$9.4 & =2.0 & $<$3.20 \\
%503-1
J033237.731-275000.41 & 503 & {\it 2.64} & 1.96 & 210 & 38.0$\pm$10.9 & 20.0$\pm$8.6 & 16.3$\pm$6.0 & 170$\pm$8  &  2.6$\pm$1.4 & SFRG & (3.5$^{+7.9}_{-2.4}$) & 46.0$\pm$8.4 & =2.0 & 2.4$\pm$0.4 \\
%%NON-DETECTIONS
{\bf \footnotesize NON-DETECTIONS} & & & & & & & & & & & & \\
%%158
J033221.624-275623.49 & 158 & ... & 1.85 & 510 & 54.9$\pm$10.9 & 28.5$\pm$8.4 & 31.2$\pm$6.0 & 38$\pm$8 & 4.8$\pm$1.4  & SMG & ... & ... & ... & ... \\
%%iv6  
J033317.754-274605.96 & {\scriptsize J033318} & ... & 2.06 & 430 & 79.9$\pm$10.8 & 72.5$\pm$8.6 & 51.4$\pm$5.9 & 100$\pm$14 & 4.3$\pm$1.4  & SFRG & ... & ... & ... & ... \\
%%iv1  
J033128.792-273916.85 & {\scriptsize J033128} & ... & ... & 460 & 105.3$\pm$11.1 & 69.6$\pm$8.7 & 39.8$\pm$6.3 & 35$\pm$8 & 4.5$\pm$1.5 & SFRG & ... & ... & ... & ...  \\
%%593  
J033248.011-275416.42 & 593 & ... & $>$2.80 & $<$30 & 18.8$\pm$11.0 & 33.5$\pm$8.6 & 12.2$\pm$6.0 & 44$\pm$8 & 9.3$\pm$1.4 & SMG & ... & ... & ... & ... \\ %2.931
\hline\hline
\end{tabular}
\end{center}
{\small {\bf Table Notes.}}  
%Thirteen 250\um\ sources observed with
%VLT ISAAC in J-, H- and K-bands, eight of which have \Ha\
%spectroscopic redshifts (six of which are secure `detections,' two are
%`tentative') and a nineth has absorption features consistent with an
%independently derived rest-UV redshift (A.M. Swinbank, private
%communication).
$^{a}$ Galaxies are split into three categories: `detections' (sources
which have reliable redshift identifications), `tentative' (poor
quality redshift identifications), or `non-detections' (no visible
emission features).  All redshift identifications are based on
\Ha\ detection except J033151.  The galaxies with `tentative'
identifications are included in all figures and tables of this paper
but are excluded from the primary analysis points in
section~\ref{sec:discussion} so as not to affect the interpretation of
this paper.

$^{b}$ ID is the identification of the 250\um\ source taken from D10
or I10.  Those from I10 are of the form J033XXX and correspond to the
first half of the BLAST name given in table 1 of I10.  Those from D10
appear as two to four digit numbers and can be found as the BLAST IDs
in table 1 of D10.

$^{c}$ Three sources have confirmed redshifts from previous
spectroscopic surveys \citep{vanzella08a,kriek08a}.

$^{d}$ $S_{350}$, $S_{500}$, and $S_{870}$ are measured directly from
BLAST/LABOCA ECDF-S maps at their respective wavelengths, using the
K-band astrometry for sources in the D10 sample, and 24\um\ and
1.4\,GHz flux densities are based on nearest neighbour matching
(described in I10).  The flux densities at 250\um, 350\um\ and 500\um\
have not been corrected for flux boosting, and their uncertainties
here only represent instrumental uncertainty; they should be combined
in quadrature with the confusion noise ($\sim$21\,mJy) for an accurate
representation of flux uncertainty.

$^{e}$ A galaxy's class is either SMG (submm galaxy) or SFRG
(submm-faint radio galaxy) based on its inclusion as a significant
detection in the \citet{weiss09b} sample, i.e. if its 870\um\ flux
density is \simgt4\,mJy.

$^{f}$ The ratio of IR luminosity to radio luminosity (as calculated
in I10 using $\alpha$\,=\,0.75, see their section~2.2), $q_{IR}$, is
given in the last column.
\end{table*}

We use the publicly released smoothed maps of the BLAST ECDF-S
\citep{devlin09a} to derive 350\um\ and 500\um\ flux densities for the
D10 sample (both 350\um\ and 500\um\ flux densities are published for
the I10 sample).  We use the K-band astrometry (where the VLT slit was
placed) as position priors to extract BLAST flux densities since the
maps are dominated by confusion noise.  We note however, that because
the maps are highly confused, they are likely to suffer from flux
boosting on individual sources, especially for the fainter
250\um\ sources; see \citet{chapin10a} for a detailed analysis of
source extraction, confusion limitations with BLAST data, and SED
fitting.

The fluxes in Table~\ref{tab:observations} are not corrected for
deboosting, but we do apply corrections when fitting FIR SEDs and
deriving $L_{\rm FIR}$, $T_{\rm dust}$ and $\beta$.  We use the
signal-to-noise ratio of the 250\um\ detections to determine the
deboosting factor which corrects for the Eddington bias and boosting
by confusion noise, as in \citet{eales09a}, figure~A2.  This results
in flux densities which are $\sim$55-75\%\ of the original measured
values for the BLAST bands given in Table~\ref{tab:observations}.  We
do not correct 870\um\ LABOCA points for deboosting since the 870\um\
flux densities are less likely to be boosted by flux from adjacent
sources.  This is because the sources which boost a high-redshift
galaxy's 250-500\um\ flux density are far more likely to sit at lower
redshifts than at higher redshifts (where the surface density of
sources is more rare).  These low redshift sources are also unlikely
to be bright at 870\um\ given typical blackbody SED shapes at low-$z$.
We recognise that our application of a deboosting factor to the BLAST
bands and not the LABOCA data is a simplification of a complex issue,
but we consider it the most realistic constraint on the FIR flux
densities given the data which are available.  We do not deboost the
{\it Spitzer} MIPS photometry at shorter wavelengths since the source
surface density at $\sim$70\um\ is similar to the source density at
\simgt200\um, yet the relative beam size is much smaller at shorter
wavelengths.

We choose not to propagate the uncertainty in the FIR deboosting
factor (which is estimated to be as large as 50\%\ at low 250\um\ S/N)
into the SED uncertainty because the deboosting factor is not
independent between bands.  Despite the large uncertainty, the FIR
flux densities for individual sources would be deboosted by similar
factors.  A correlated deboosting factor between bands would imply
less uncertainty in derived $T_{\rm dust}$ or $\beta$ than blindly
adopting the deboost uncertainties.  We test the correlation of the
deboosting factor by using the FIR colours ($S_{250}$/$S_{350}$ and
$S_{250}$/$S_{500}$) of {\it Herschel} SPIRE and PACS sources
\citep{amblard10a}.  Using Monte Carlo tests, we remove the
contribution of a single potential boosting source by subtraction of
an arbitrary 250\um\ flux and the associated 350\um\ and
500\um\ fluxes associated with the colours of a randomly selected
galaxy from the \citet{amblard10a} sample.  We find that the FIR
luminosity does not vary by more than $\pm$0.1\,dex and that dust
temperature varies by about $\pm$9\,K.  If two contaminating boosting
sources are incorporated with different FIR colour properties, the variance on
the fitted $L_{\rm FIR}$ and $T_{\rm dust}$ decreases further.  We
discuss the impact that the deboosting factor uncertainty has on our
final conclusions more in \S~\ref{sec:confusion}.
%While these uncertainties are quite large
%and vary greatly from source to source, the deboosting factor is not
%independent between the different bands, thus it would not impact the
%derived $T_{\rm dust}$ or $\beta$ nearly as significantly as the FIR
%luminosities.  In other words, the flux densities would move up or
%down together by similar factors (proportional to beamsize which does
%not change greatly between FIR bands).
%
%It should be kept in mind that the uncertainty in
%deboosting introduces another uncertainty into our derived quantities
%and SED fits which is not represented in this paper, but should be
%investigated carefully in the future.  

We measure 870\um\ flux densities (at the K-band positions) from the
LABOCA map of ECDF-S \citep{weiss09b}.  Six sources have
$S_{870}$\simgt\,4\,mJy and are listed in
\citet{weiss09b}.  J033246 is claimed as an SMG in D10, but its
870\um-peak is $>$30\arcsec\ away from its K-band position.  This
implies that 7/13 ($\sim$53\%) of our sample are submm-faint and would
be excluded from traditional SMG surveys.  All galaxies except J033212
are also radio detected in VLA data at $>$30\,\uJy.

The galaxies' rest-frame near-IR photometry is consistent with stellar
emission, from which we derive stellar masses from rest-frame $H$-band
(Table~\ref{tab:halpha}), using the methods described by
\citet{hainline09a}.  Measuring the absolute magnitude of a galaxy
near its 1.6\um\ ``stellar bump'' provides the most accurate measure
of its stellar mass, however it is reliant on the assumption of a
constant mass to light ratio (here we assume M/L\,=\,3.2), reddening
properties, and minimal AGN contribution to near-IR flux.  The stellar
mass estimates are uncertain by $\sim$0.3dex.  The near-IR photometry
is also be used to infer AGN content, since a flux excess at 8\um\
(significantly above stellar population model fits) is indicative of power
law emission from an AGN.  None of our sources have $>$2$\sigma$ 8\um\ flux
excesses.

Two of the 13 observed sources, J033151 and J033152, are X-ray
detected above the luminosities which would correspond with their star
formation rates (L$_X$\,\simgt\,10$^{44}$\,erg\,s$^{-1}$).  Only one
individual source shows obvious signs of containing a luminous AGN,
from its radio flux excess and detection in the X-rays: J033152.  From
this AGN estimator (and the analysis in the ensuing section about
\Ha\ properties), we infer roughly that 20$\pm$15\%\ of 250\um-bright
sources have signs of dominant AGN.
%  We use the ratio of
%  N~II to H$\alpha$, X-ray imaging, and powerlaw fit to near-infrared
%  photometry to show that Active Galactic Nuclei (AGN) contribute to
%  \simlt\,20\%\ of the bolometric output of these Ultraluminous
%  Infrared Galaxies (ULIRGs).  

\subsection{\Ha\ Properties}

\begin{figure}
\centering
\includegraphics[width=0.99\columnwidth]{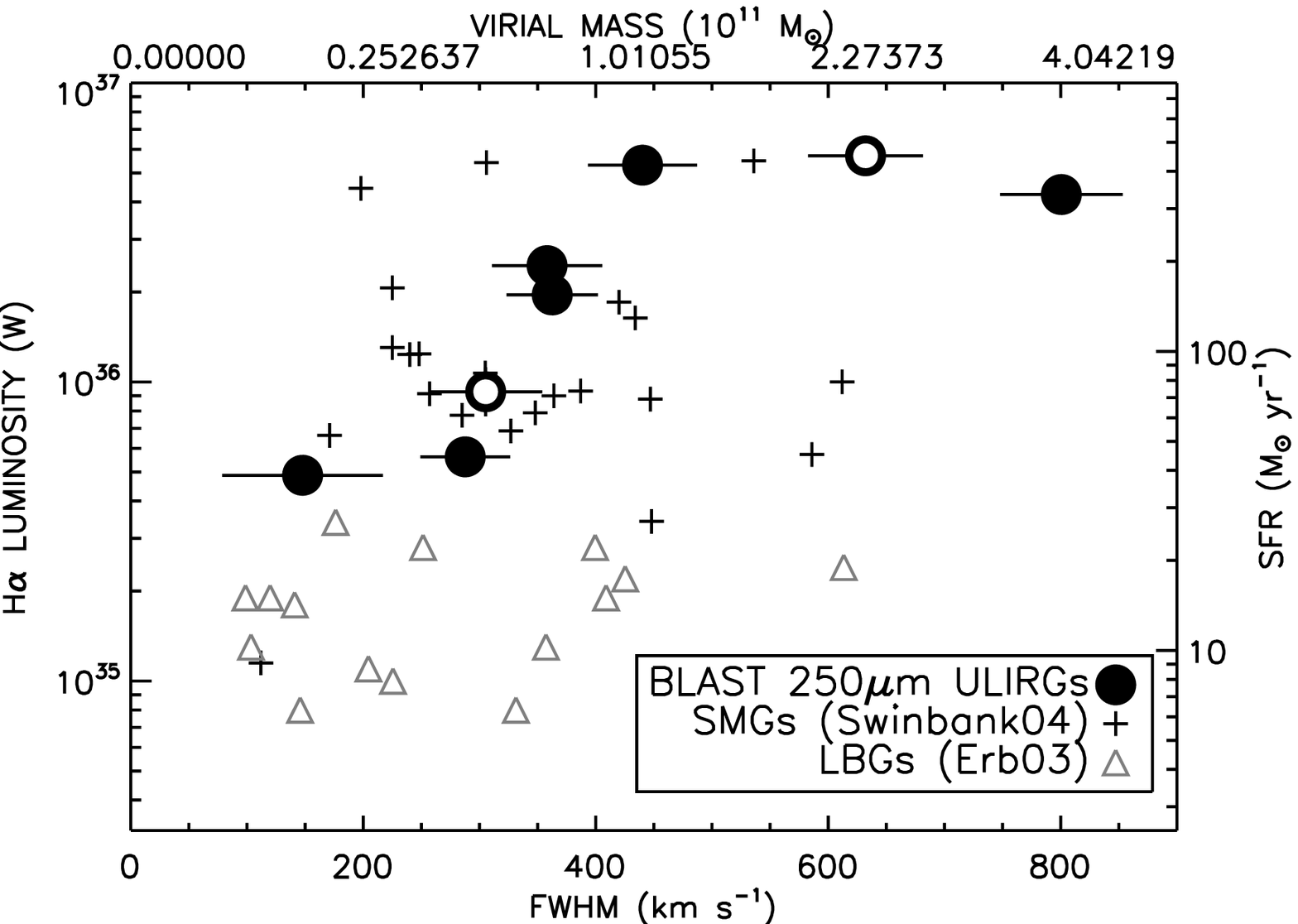}
\caption{ The relation between \Ha\ FWHM and line luminosity is shown
  below, for the BLAST sample, SMGs \citep[crosses;][]{swinbank04a},
  and normal z$\sim$2 galaxies \citep[e.g. LBGs, gray
  triangles;][]{erb03a}.  The \Ha\ line luminosity is converted to a
  SFR at right and the \Ha\ FWHM is converted to a virial mass
  estimate (using the assumption of $r_{1/2}$\,=\,3\,kpc) at top.  The
  BLAST 250\um\ sample has similarly broad and strong \Ha\ emission to
  the SMG population, and is brighter than the normal z$\sim$2
  galaxies.  The two galaxies with tentative redshifts are
  shown as empty circles.  This plot suggests that 250\um-luminous
  galaxies might share intrinsic, physical properties with the SMG
  population.  }
\label{fig:stackedha}
\end{figure}

We measure \Ha\ line widths and \Nii/\Ha\ ratios in order to infer AGN
content from the six BLAST sources for which we have secure
\Ha\ observations (i.e. not including J033151).  After deconvolving
the measured full width half maxima (FWHM) with the instrumental
resolution measured from skylines in the vicinity of
\Ha\ ($\sim$6.5\AA\ in K-band and $\sim$4.4\AA\ in H-band) we find
that our \Ha\ lines have an average rest-frame FWHM of
415$\pm$20\,\kms\ and span the range 150-800\,\kms\ (for the six
galaxies with secure \Ha\ detections).  The individual \Ha\ properties
of each galaxy are given in Table~\ref{tab:halpha}.  We plot the
\Ha\ FWHM against \Ha\ luminosity in Figure~\ref{fig:stackedha}.  The
\Ha-inferred star formation rates average to 210$\pm$160\,\Mpy, which
requires a substantial extinction factor to account for the star
formation observed in the FIR (on the order of 2000\,\Mpy).  The mean
SFR$_{\rm FIR}$/SFR$_{H\alpha}$ ratio for the sample is 36$\pm$22,
which is comparable to the ratio for the SMG population of
\citet{swinbank04a} of 31$\pm$15. The subset of our sample which is
submm-faint (SFRGs) also have similar SFR ratios, averaging 37$\pm$24.

For the six galaxies which have secure \Ha\ observations, a stacked
spectrum is shown in Figure~\ref{fig:isaacspectra} which we use to
measure the aggregate line emission properties of the sample.  The
\Ha\ line width of the stacked spectrum is
530$\pm$280\,\kms\ (statistically indistinguishable from the
individual \Ha\ measurements or the mean SMG line width, 390\,\kms),
and a line luminosity corresponding to a star formation rate of
190\,\Mpy.  The line width is slightly larger than the mean line width
for the sample likely due to signal-to-noise limitations of the
original data.  Both line width measurements, 415\,\kms\ and
530\,\kms, are consistent with the dynamics of active star forming HII
regions, except the high FWHM outlier: J033243 at z\,=\,2.123 with
FWHM\,=\,800\,\kms. J033243 is also the second brightest \Ha\ emitter
with a high \Ha\ implied SFR, 335\,\Mpy; its SFR$_{\rm
  FIR}$/SFR$_{H\alpha}$\,=\,7, which is the lowest SFR ratio of the
sample indicative of a less \Ha\ obscuration.

We convert the \Nii/\Ha\ ratios to $12+\log({\rm O/H})$ using the
methods described by \citet{maiolino08a}. However two sources (J033249
and J033212) have O/H limits $>$9.25, which corresponds to very strong
\Nii/\Ha (\simgt\,0.5).  The O/H and \Nii/\Ha\ indicators saturate at
metallicities above solar \citep[see][]{pettini04a}, and additional
contribution from either AGN or shocked gas can increase the \Nii/\Ha\
ratio further \citep[e.g.][]{vandokkum05a}. The remaining five secure
detections have $12+\log({\rm O/H})$ values which average to
8.64$\pm$0.08.  The measured \Nii/\Ha\ ratio for the stacked \Ha\
spectrum implies a metallicity of $12+\log({\rm
O/H})$\,=\,$8.75_{-0.10}^{+0.08}$ (in agreement with the average for
the individual measurements).

\begin{table*}
\begin{center}
\caption{\Ha\ and stellar properties of BLAST 250\,\um\ Galaxies}
\begin{tabular}{ccccccccccc}
\hline\hline
NAME & $z$ & $S_{\rm H\alpha}$ & $L_{\rm H\alpha}$ & $FWHM_{\rm H\alpha}$ & $\langle{\rm [NII]/H\alpha}\rangle^a$ & $\langle{\rm O/H}\rangle^a$ & SFR$_{\rm H\alpha}$ & SFR$_{\rm FIR}$ & SFR$_{\rm radio}$ & $M_{\star}$ \\
 & & (W\,m$^{-2}$) & (10$^{35}$\,W) & (\kms) & & & (\Mpy) & (\Mpy) & (\Mpy) & (\msun) \\
\hline
J033129 & 1.482 & 4.1$\times$10$^{-19}$ & (5.6$_{-0.8}^{+0.9}$) & 190$\pm$60 & 0.13 & 8.60 & 44$^{+7}_{-6}$   & 2100$^{+1100}_{-700}$     & 1300$^{+440}_{-330}$ & 1$\times$10$^{10}$ \\
J033151 & 1.599 & ...                   & ...                   &        ... & ... & ...   &     ...          & 3800$^{+2500}_{-1500}$     & 1000$^{+430}_{-300}$ & 3$\times$10$^{10}$ \\
J033152 & 2.342 & 1.3$\times$10$^{-18}$ & (53$_{-2}^{+3}$)     & 490$\pm$40 & 0.74 & $>$9.25 & 419$^{+20}_{-19}$ & 14000$^{+10000}_{-6000}$ & 28000$^{+5200}_{-4400}$ & 3$\times$10$^{10}$ \\
J033204 & 2.252 & 5.2$\times$10$^{-19}$ & (20$_{-3}^{+3}$)      & 360$\pm$40 & 0.14 & 8.64 & 154$^{+24}_{-21}$ & 6800$^{+1700}_{-1400}$    & 3300$^{+1000}_{-800}$ & 5$\times$10$^{10}$ \\
J033243 & 2.123 & 1.3$\times$10$^{-18}$ & (42$_{-5}^{+5}$)      & 800$\pm$50 & 0.18 & 8.71 & 335$^{+46}_{-40}$ & 2400$^{+4300}_{-1500}$    & 2000$^{+720}_{-530}$ & 4$\times$10$^{10}$ \\
J033246 & 1.382 & 4.2$\times$10$^{-19}$ & (4.9$_{-0.5}^{+0.5}$) & 150$\pm$70 & 0.13 & 8.59 &  39$^{+4}_{-4}$   & 680$^{+200}_{-150}$     & 650$^{+180}_{-140}$ & 1$\times$10$^{10}$ \\
J033249 & 2.326 & 5.8$\times$10$^{-19}$ & (25$_{-3}^{+3}$)      & 360$\pm$50 & 0.14 & 8.64 & 194$^{+23}_{-20}$ & 14000$^{+7000}_{-4000}$   & 6100$^{+1700}_{-1300}$ & 3$\times$10$^{10}$ \\
\hline
J033212 & 1.93 & 3.5$\times$10$^{-19}$ & (9.3$_{-1.6}^{+1.9}$) & 310$\pm$50 & 0.55 & $>$9.25 & 73$^{+15}_{-13}$ & 3800$^{+7000}_{-2000}$   & $<$690 & 4$\times$10$^{10}$ \\
J033237 & 2.64 & 1.0$\times$10$^{-18}$ & (57$_{-8}^{+9}$)      & 630$\pm$50 & 0.33 & 8.96 & 451$^{+69}_{-60}$ & 6000$^{+13000}_{-4000}$     & 6700$^{+1500}_{-1200}$ & 9$\times$10$^{10}$ \\
%J033221... & 2.277 & (2.5$_{-0.4}^{+0.5}$)$\times$10$^{36}$ & 563$\pm$48 & 197$^{+37}_{-31}$ \\
\hline\hline
\label{tab:halpha}
\end{tabular}
\end{center}
{\small {\bf Table Notes.}  \Ha\ properties from VLT-ISAAC spectra of
  the BLAST 250\um\ sample.  The seven at top have secure redshifts
  (the six which have \Ha\ properties are used in our analysis) while
  the bottom two have tentative redshifts (we calculate their \Ha\
  properties, but exclude them from aggregate property analysis in
  section \ref{sec:discussion} despite being illustrated in figures).
  FWHM has been deconvolved with the instrumental resolution measured
  from skylines in the vicinity of \Ha\ ($\sim$6.5\AA\ in K-band and
  $\sim$4.4\AA\ in H-band), and SFR is derived from $L_{\rm H\alpha}$
  using the relation $SFR\,=\,7.9\times10^{-35}L_{H\alpha}$ from
  \citet{kennicutt98a}. The metallicity measurements ($^a$) are
  computed by $\langle{\rm NII/H\alpha}\rangle$\,=\,$\log({\rm
  S_{NII}/S_{H\alpha}})$ and $\langle{\rm
  O/H}\rangle$\,=\,12+$\log({\rm O/H})$ derived from $\langle{\rm
  NII/H\alpha}\rangle$ using methods described in \citet{maiolino08a}.
  The characteristic uncertainty on \Nii/\Ha\ is $\sim$0.10.  Stellar
  masses ($^b$) are measured from $Spitzer$-IRAC photometry which
  brackets the rest-1.6\um\ stellar bump (see \S~2); the
  characteristic uncertainty in stellar mass is
  $\sim$2$\times$10$^{10}$\,\msun. }
\end{table*}

\subsection{Dust SED Fitting and FIR/radio Correlation}\label{sec:dustsed}

We fit the MIPS (70\um, and 160\um, where available), BLAST (250\um,
350\um, and 500\um), and LABOCA (870\um) flux densities to two
different FIR dust models.  For FIR SED fitting, we correct the BLAST
flux densities for boosting by confusion noise as mentioned in the
beginning of this section.  Both FIR SED models assume a modified
blackbody emission curve with a single dust temperature:
\begin{equation}
S_{\nu}\,\propto\,\frac{\nu^{3+\beta}}{exp(h\nu/kT_{dust})-1}
\label{eq:blackbody}
\end{equation}
where $S_{\nu}$, the flux density, is a function of rest frequency
$\nu$, the emissivity $\beta$, dust temperature $T_{d}$, and FIR
luminosity $L_{FIR}$ (which governs the normalisation of the
function).  The first model allows $\beta$ to vary (the ``beta-free''
model) while the second model fixes emissivity to $\beta\,=\,2$.  Both
models have $T_{d}$ and $L_{FIR}$ as free parameters.  The advantage
of allowing emissivity to vary in the first model allows a
reassessment of the emissivity constraints which have been placed on
ULIRGs in past studies \citep[e.g. $\beta\,=\,$1.5 or
2.0;][]{chapman05a,casey09b,casey09c,younger09a}.  In addition, our
measurements of $\beta$ are made independent of any $a\ priori$
constraint on $T_{d}$ or $L_{FIR}$.  We choose to make the second
model rigid as fits from the first model can be unphysical, as might
be the case if the FIR flux densities are particularly faint or
affected by source confusion.

\begin{figure*}
\centering
\includegraphics[width=1.99\columnwidth]{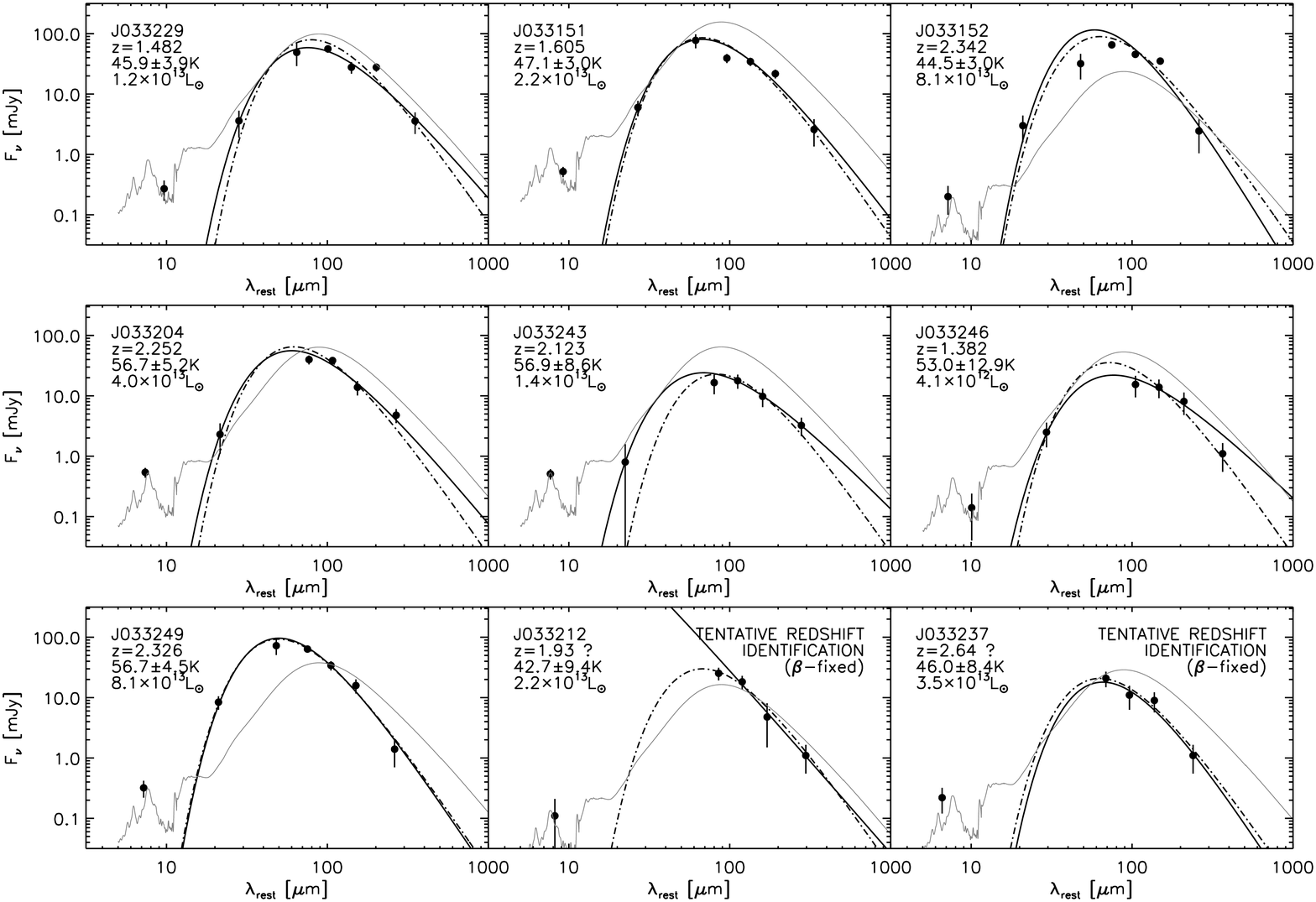}
\caption{ The FIR SEDs for high-$z$ BLAST sources, including data from
  $Spitzer$-MIPS (24\um, and 70\um, 160\um\ where available), BLAST
  (250\um, 350\um, and 500\um) and LABOCA (870\um).  We fit three SEDs
  to the data: (1) a single dust temperature with fixed $\beta\,=\,2$
  modified blackbody as in Eq.~\ref{eq:blackbody} (dashed-dotted
  line), (2) a modified blackbody with $\beta$ treated as a free
  parameter (solid black line), and (3) an SMG composite SED from
  \citet{pope08a} normalised to 24\um\ flux density (thin gray line).
  The galaxies' names, redshifts, best fit dust temperatures, and FIR
  luminosities are inset on each SED plot.  The two galaxies marked
  {\sc tentative} (from their tentative redshift identifications) are
  also the two galaxies whose fixed $\beta\,=\,2$ SED fit was significantly
  better than the beta-free model.  Neither tentative galaxies are included in the analysis of section~\ref{sec:discussion}}
\label{fig:firsed}
\end{figure*}

Only J033212 and J033237 are poorly fit to a beta-free model (these
are the two galaxies with tentative redshift identifications), since
they do not have 70\um\ data and have unconstraining upper limits in
the FIR. We use only the fixed $\beta$\,=\,2 model for these two.  The
remaining seven galaxies have reliable beta-free SED fits, and from
them we measure $\beta$, $T_{dust}$, and $L_{FIR (8-1000\mu m)}$
(summarised in Table~\ref{tab:observations}).  Both fixed beta and
beta-free fits are shown in Figure~\ref{fig:firsed}.  We find a mean
emissivity of $\beta$\,=\,1.73$\pm$0.13 and a mean dust temperature of
$T_{d}$\,=\,52$\pm$6\,K.

The FIR luminosities (8-1000\um) must be corrected to account for
mid-infrared (8-25\um) emission from PAH and power law sources
\citep[e.g.][]{menendezdelmestre09a} above the single FIR modified
blackbody.  We tether the \citet{pope08a} SMG SED to 24\um\ flux
densities (as seen in Figure~\ref{fig:firsed}) to estimate the
luminosity deficit of the single temperature blackbody.  This deficit
varies substantially object to object due to the large spread in
24\um\ flux densities and blackbody properties in the 8-25\um\ Wein
tail.  On average, we find that the contribution of the PAH and AGN
emission account for 0.04$\pm$0.03\,dex of luminosity which we add to
the FIR luminosities as a correction factor.  Although the mid-IR
properties of the sample can vary substantially, this deficit
translates to no more than a $\sim$10\%\ increase in FIR luminosity
for these $>$10$^{13}$\,\lsun\ systems.  The corrected luminosities
are given in Table~\ref{tab:observations}.

%A correction factor must be applied to the FIR luminosities measured
%directly from the single temperature blackbody fits between 8-1000\um.
%Towards shortest wavelengths ($\sim$8-25\um), the FIR blackbody
%emission from the best-fit SEDs becomes negligibly small
%($\ll$\,0.1mJy) compared to the observed 24\um\ flux densities
%($\sim$0.3-0.7mJy) and the suspected powerlaw and PAH emission which
%dominates the mid-IR emission in dusty starburst galaxies
%\citep{menendezdelmestre09a}.  
%The correction factor for $L_{\rm FIR}$
%accounts for the additional mid-IR flux above the FIR blackbody in the
%rest-frame $\sim$8-25\um\ wavelength range.  Since this wavelength
%range is poorly covered in our BLAST sample, we use the
%\citet{pope08a} composite mid-IR SMG SED, normalised to 24\um\ flux
%(as shown in Fig.~\ref{fig:firsed}), to calculate the 8-25\um\
%luminosity excess, $C$, above the best-fit blackbody SED.  In other
%words, C\,$\equiv$\,$L_{FIR (8-25\mu
%m)}$[$\langle$SMG$\rangle$]-$L_{FIR (8-25\mu m}$[best-fit blackbody].
%We calculate that a mean luminosity correction factor of
%C\,=\,0.044$\pm$0.032\,dex (on order $\sim$10\% increase in $L_{\rm
%FIR}$) is needed to account for the underestimated FIR SED luminosity.
%The luminosities in table~\ref{tab:observations} have had the
%correction applied, and the average for the whole sample is
%$L_{FIR}$\,=\,3$\times$10$^{13}$\,\lsun.

We also overplot the composite SMG spectrum, from \citet{pope08a},
normalised to the integrated 24\um\ flux density in
Figure~\ref{fig:firsed}.  While the SMG composite is carefully derived
based on mid-IR to FIR data of SMGs to date, it fails to fit the BLAST
FIR data on a case by case basis.  In some cases, it
under/overestimates the FIR luminosities by $\pm$1\,dex.  This
illustrates how a 24\um-normalised SED fitting procedure, which is
common in the literature \citep[e.g.][]{desai09a} places poor
constraints on the breadth of FIR properties of ULIRG samples,
especially in the absence of direct FIR measurements.  Recent
high-resolution FIR observations \citep[e.g.][]{younger10a} have
demonstrated that 24\um\ counterparts are often misidentifications and
do not correspond to the FIR luminous source.

Although multiple dust temperature blackbodies are found to fit well
to local ULIRGs in the literature \citep[for example,
  see][]{clements09a}, strong assumptions must be made regarding the
FIR luminosity or normalisation, to decompose the sparse FIR data down
into multiple blackbody components.  Given the uncertainty of the FIR
luminosities or flux densities at any given wavelength, we decide to
forgo multiple dust temperature fitting for well constrained, single
dust temperature blackbody fits. If multiple blackbodies provide a
more physical SED fit, then our derived emissivities, from the single
blackbody fits, could be underestimated.

We measure $q_{\rm IR}$, the ratio of integrated IR flux to
radio flux, as described in detail by I10 to verify the FIR/radio
correlation in our sample.  I10 finds a mean $q_{\rm
IR}$\,=\,2.41$\pm$0.20 based on the larger sample of BLAST sources
with photometric redshifts.  Assuming a radio synchrotron slope of
$\alpha$=0.75, we measure a mean $q_{\rm IR}$\,=\,2.46$\pm$0.18 which
agrees with I10 and earlier findings \citep[e.g.][]{dale07a} that
there is no evidence for evolution in $q_{\rm IR}$ with redshift.
Since $\alpha$ has a significant impact on the calculation of $q_{\rm
IR}$, we consider the impact of variations in $\alpha$: I10 explicitly
measured $\alpha$ for a high-redshift subset of their BLAST sample and
found a median value of $\alpha$\,=\,0.4.  If we use $\alpha$\,=\,0.4
instead to calculate $q_{\rm IR}$ we measure $q_{\rm
IR}$\,=\,2.27$\pm$0.17.  This is still in agreement with I10's
measurement of the FIR-radio correlation at high redshift within
uncertainties.

\begin{figure}
\centering
\includegraphics[width=0.99\columnwidth]{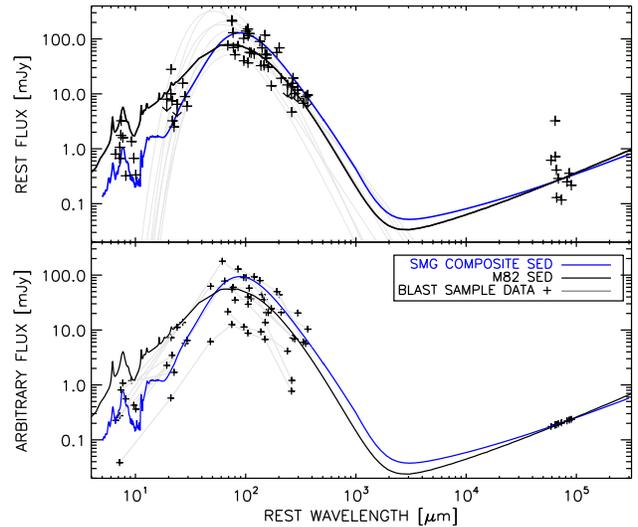}
\caption{ A comparison of the BLAST FIR and radio flux densities
  against two template SEDs: an M82 spectrum (black) from
  \citep{bressan02a} and a composite SMG spectrum (blue) taken from
  \citet{pope08a}.  The upper panel plots all sources' flux densities
  in their rest-frame (with overplotted best-fit FIR SEDs (light gray)
  from Fig.~\ref{fig:firsed}).  The lower panel plots all sources with
  fluxes renormalised to their radio points.  The scatter around the
  template SEDs in the lower panel (where the flux has been
  renormalised to the radio) indicates our observed variation in the
  FIR-radio correlation.  While there appears to be a large scatter,
  there is no systematic offset from the FIR-radio predicted FIR
  luminosities and those measured directly in our sample. }
\label{fig:firradio}
\end{figure}

We note that \citet{kovacs06a} concluded that the local FIR/radio
correlation {\it overestimates} FIR luminosity by factors of
$\sim$0.2-0.4\,dex for SMGs, which contrasts with our and I10's
result for BLAST sources.  However, the difference is due to different
FIR SED fitting procedures; when we refit the 21 SMG FIR points
(measured at 350\um\ and 850\um) in \citet{kovacs06a} using the
methods described in this paper (for fixed $\beta$\,=\,1.5) we find
$q_{\rm IR}$\,=\,2.46$\pm$0.19. Figure~\ref{fig:firradio} shows the
flux densities of the BLAST sample in the FIR-radio context, plotted
against two template SEDs which follow the relation: an M82 template \citep{bressan02a}
and a composite SMG SED \citep{pope08a}.  Some sources exceed the
relation (often due to flux excesses towards shorter wavelengths)
while others' FIR luminosities are over-predicted, most likely due to
AGN contribution to radio luminosity (e.g. J033152).  However, these
variations do not appear to be weighted in either direction,
indicating that the FIR/radio correlation is predicting FIR
luminosities accurately for the population on a whole.

%In the absence of constraining FIR data (as was the case for SMGs and
%other high-z ULIRGs), the FIR/radio correlation has been used to
%derive FIR luminosities using radio flux density ($S_{1.4GHz}$, in
%\uJy), radio spectral slope ($\alpha$=0.75) and redshift:
%\begin{equation}
%\label{eq:firradio}
%L_{FIR} = 2.21\times10^{-48}\,S_{1.4GHz}\,(1+z)^{(1-\alpha)}\,4\pi\,d_{L}^{2}
%\end{equation}
%At z$\ll$1, this relation infers a value of $q_{\rm IR}$\,=\,2.35,
%however at z$\sim$2, $q_{\rm IR}$\,=\,2.12 is implied and at z$\sim$3,
%this drops to $q_{\rm IR}$\,=\,2.06.  Although there is no observed
%evolution in $q_{\rm IR}$ with redshift (and no suggested evolution in
%the FIR/radio correlation), the behaviour of Eq.~\ref{eq:firradio}
%predicts FIR luminosities for this sample ranging from
%2$\times$10$^{12}$-6$\times$10$^{13}$\,L$_{\odot}$.  This is, on
%average, $\sim$0.5dex below the FIR luminosity range given in
%Table~\ref{tab:observations}.  The difference is largely due to the
%assumption of $\alpha$\,=\,0.75 for use in the K-correction; the value
%of $\alpha$ can impact the derived FIR luminosity much more
%significantly at higher redshift. 
%A value of
%$\alpha$\,$\approx$\,1.2 is needed to recover the measured $q_{\rm
%IR}\sim$2.4.  Although the uncertainties in $q_{\rm IR}$ and $\alpha$
%are large, it is clear that our measurements are inconsistent with the
%FIR/radio correlation as stated in Eq.~\ref{eq:firradio}, which
%underestimates FIR luminosities by $\sim$0.5\,dex.

\section{Discussion}\label{sec:discussion}

Combining the unique BLAST FIR data with previous \textit{Spitzer}
observations and our spectroscopic redshift identifications allows a
full characterisation of FIR SEDs for 250\um-bright HyLIRGs.  Since
250\um\ mapping has the advantage of sampling blackbody emission near
its peak at z$\sim$2, 250\um\ ULIRG/HyLIRG selection is far less
biased towards certain SED shapes, making 250\um-bright galaxies
a much cleaner, unbiased subset all high-redshift ULIRGs.

\subsection{Source Density}\label{sec:sourcedensity}

The seven galaxies with secure redshifts suggest that these
250m-bright galaxies have roughly 1/5 the volume density of similarly
luminous SMGs.  We exclude the sources which only have tentative
redshift identifications from this analysis, although we note that
they would not skew or affect the interpretations we draw from the
entire sample.  We estimate this lower limit to $\rho_{250}$ for our
sample by taking the redshift ranges $z=1.3-1.7$ and $z=2.0-2.6$,
given the gap in the atmospheric opacity between $H$ and $K$-bands.
We treat the D10 and I10 sample separately as the relative selection
depths differ ($S_{250}>$33\,mJy and $S_{250}>$59\,mJy respectively),
finding $>8\times10^{-6}$\,h$^3$\,Mpc$^{-3}$ and
$>5\times10^{-6}$\,h$^3$\,Mpc$^{-3}$ respectively.  Of the
spectroscopically identified SMG samples in the literature
\citep[e.g.][]{chapman05a}, 51\%\ of the sources have L$_{FIR} >
8\times10^{12}$\,\lsun (a cutoff corresponding approximately to the
BLAST 250\um\ depth), which implies a luminosity-limited volume
density of $2.5\times10^{-5}$\,h$^3$\,Mpc$^{-3}$ for SMGs.  As much
deeper, more uniform 250\um\ data become available from {\it
Herschel}, the overlap with the SMG population is being explored more
fully \citep[e.g.][]{elbaz10a}.

Of our nine spectroscopic sources, only four have been detected as
SMGs in \citet{weiss09b}.  While 250\um-bright sources at high-$z$ are
more rare than SMGs, the fact that 55\%\ (5/9) of our sample are
submm-faint (with $\langle S_{\rm 870}\rangle$\,\simlt\,2\,mJy)
highlights that the SMG population represents only a subset of
high-redshift ULIRG activity, as \citet{casey09a}, \citet{chapman04a}
and \citet{blain04a} suggest.  The addition of 250\um-selected,
submm-faint galaxies to the previously-known HyLIRG population could
imply that the volume density of known high-$z$ HyLIRGs would increase
from the SMG estimate roughly by \simgt\,12\%.  However, more spectral
observations of similar 250\um-bright objects from \textit{Herschel}
are needed to boost these statistics and understand the actual level
of contribution.
%Further evidence that 250\um\ and
%850\um-bright populations would be distinct is in our calculation of
%blackbody emissivity; on average, we find $\beta\,$=\,2.5$\pm$0.5,
%which is much steeper than $\beta\,=\,$1.5, the nominal value used to fit
%ULIRG FIR SEDs.  Greater emissivity translates to a steeper
%Rayleigh-Jeans tail and a larger disparity between 250 and
%850\um\ flux densities, so that a 250\um-bright source is less likely
%to be detected at 850\um.  An increased dependence on $T_{dust}$ is also
%introduced into 850\um\ samples with larger $\beta$.

\subsection{Near-infrared Spectral Features}

The \Ha\ derived star formation rates of the BLAST HyLIRG sample
underestimate the FIR SFRs by $\sim$10$\times$, as is often the case
with rest-UV or rest-optical emission line star-formation indicators
in dust-obscured starburst galaxies.  However, we note near-IR
spectroscopic observations of SMGs in \citet{takata06a} measured
internal extinction factors of $A_{V}$\,=\,2.9$\pm$0.5 using
\Ha/H$\beta$ ratios.  When correcting the \Ha-inferred SFRs in
Table~\ref{tab:halpha} for this dust extinction the FIR-inferred SFRs
are recovered, averaging to $\sim$2000\,\msun\,yr$^{-1}$.  This
indicates that dust obscuration is significant in the near-IR and must
be corrected for to understand the true nature of the ultraluminous
activity in these galaxies.

Placing our \Nii/\Ha\ metallicity measurements in a larger galaxy
evolution context, the metallicities of this sample (measured by
converting to $\langle{\rm O/H}\rangle$,
i.e. $\langle12\,+\,\log($O/H$)\rangle$\,=\,8.65), agree within
uncertainties with the observed metallicities of the most massive
z$\sim$2 galaxies,
$\langle12\,+\,\log($O/H$)\rangle\,\sim$\,8.55$\pm$0.07, in
\citet{erb06a}.
%While the stellar masses of our sample lie in the
%10$^{10}$-10$^{11}$\msun\ range instead of $>$10$^{11}$\msun, we note
%that the stellar mass estimates of z$\sim$2 normal galaxies might be
%overestimated due to improper accounting for the AGN contribution to
%mid-IR flux.  High-$z$ ULIRGs have been shown to have consistently
%high stellar masses with respect to other similar epoch galaxy
%populations, so it is thought that
The mean \Nii/\Ha\ ratio for this sample, 0.29$\pm$0.23, agrees within
uncertainty with the \citet{swinbank04a} SMGs, 0.41$\pm$0.38.  While
evolutionary conclusions should not be drawn from these data alone,
the results are consistent with conjecture that the ULIRG phenomenon
occurs at the early stages of a burst in star formation triggered by
the merger of two typical gas-rich massive galaxies at z$\sim$2.

%dynamical masses from Halpha relative to SMGs that have >10^13 Lsun.

\subsection{Temperature Fitting and Selection}

\begin{figure*}
\centering
\includegraphics[width=1.8\columnwidth]{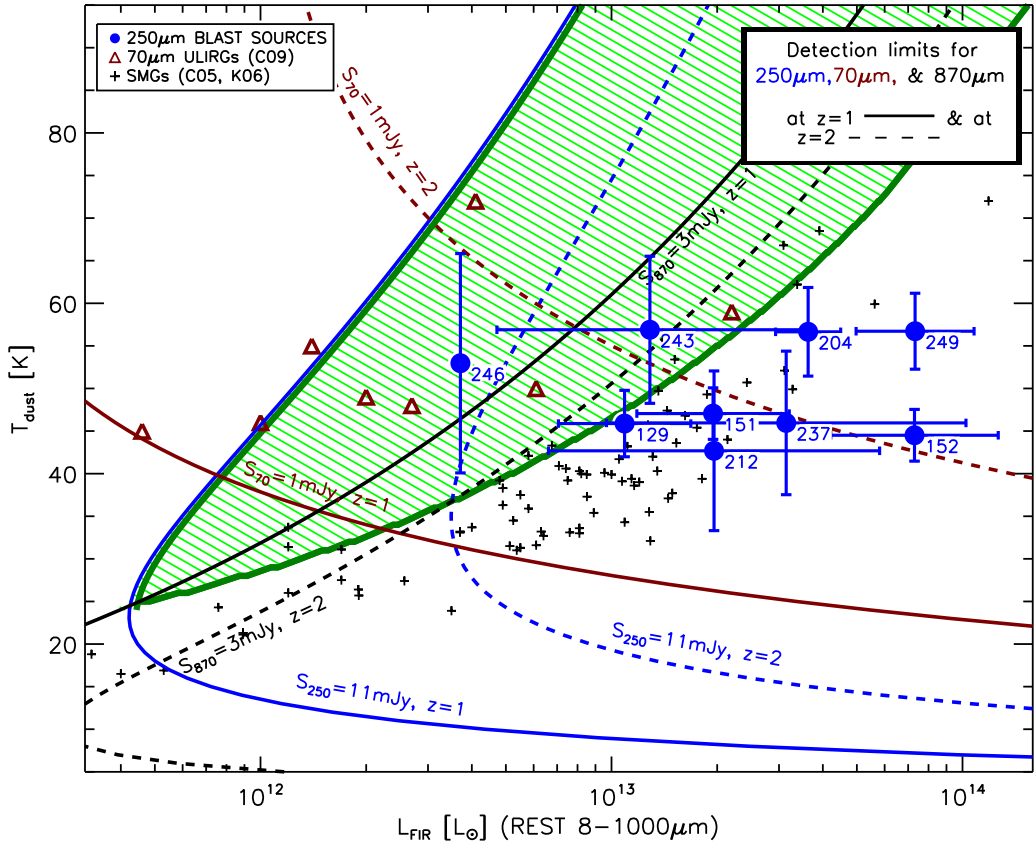}
\caption{ FIR luminosity against dust temperature for the BLAST
  250\um\ sample (blue circles, labelled by the last three digits in
  their right ascension).  The dashed ($z\,=\,2$) and dotted
  ($z\,=\,1$) lines indicate rough 1-$\sigma$ boundaries at
  250\um\ (blue, $\sigma_{250}$=11mJy, BLAST), 70\um\ (red,
  $\sigma_{70}$=1mJy, MIPS), and 870\um\ (black, $\sigma_{870}$=3mJy,
  LABOCA), where sources at the given redshift would have 1-$\sigma$
  significance if it has $L_{\rm FIR}$ and T$_{d}$ corresponding to
  the boundary, and would be $>$1$\sigma$ if it lies to the right of
  the boundary.  To translate these curves into 3-$\sigma$ detection
  limits, they would be shifted $\sim$0.4\,dex to the right in
  luminosity space; the shape of the boundary curves would be
  maintained.  70\um-bright ``hot-dust'' ULIRGs from \citet{casey09b}
  are overlaid as red triangles.  The 850\um-detected SMGs from
  \citet{chapman05a} and \citet{kovacs06a} are overlaid as small black
  crosses.  The area enclosed by the green highlights a phase-space
  where 250\um\ observations are more sensitive than 870\um\ at all
  redshifts $z\,>$\,1.}
\label{fig:lfirtd}
\end{figure*}

Figure~\ref{fig:lfirtd} shows dust temperature ($T_{dust}$) against
FIR luminosity, with BLAST 250\um\ sources and SMGs overplotted.
Representative 1$\sigma$ detection boundaries at 70\um, 250\um\ and
870\um\ are shown to illustrate the populations' selection biases
(2$\sigma$, 3$\sigma$ or 5$\sigma$ detection limits would have the
same shape but be shifted to the right in luminosity; e.g. the
3$\sigma$ detection limit corresponds to a luminosity shift of
$\sim$0.4\,dex).  The mean dust temperature of our sample, when fit
with single modified blackbody SEDs, is 52$\pm$6\,K, which is
comparable to the mean dust temperature of local ULIRGs of similar
(\simgt\,10$^{13}$\,\lsun) luminosities, 45$\pm$10\,K
\citep{chapman03a,rieke09a}, and only $\sim$5\,K warmer than SMGs of
similar luminosities $>$10$^{13}$\,\lsun (and is 15\,K warmer than
SMGs on average, which are 36$\pm$7\,K).  Overall, all BLAST sources
(except the lower-redshift J033246) have dust temperatures consistent
with the high luminosity end of the SMG distribution.

It is important to note that the SMG FIR luminosities shown here are
derived from radio luminosity, via the FIR/radio correlation, and that
the associated dust temperature fits are reliant upon that assumption.
To first order, we and others have shown that the FIR/radio
correlation holds at these redshifts and luminosities (for direct
comparison see Table~\ref{tab:halpha}), however scatter is
significant, with differences in measured/derived FIR luminosities of
$\pm$0.7\,dex.  As is often done for literature SMGs to date
\citep{chapman05a}, dust temperature is measured by using a single FIR
data point (e.g. observed 850\um) and forcing an SED with fixed
radio-inferred $L_{\rm FIR}$.  If the radio-inferred FIR luminosity is
significantly different from the actual FIR luminosity, then the dust
temperature will either be grossly over- or underestimated.

We use the BLAST sample and its full SED information (thus directly
measuring FIR luminosity) to test the accuracy of FIR fits and derived
dust temperature for SMGs and other high-$z$ ULIRGs.  Regardless of
the accuracy of radio-derived FIR luminosities, we find that $T_{\rm
dust}$ is systematically underestimated by 12$\pm$19\,K when derived
from 870\um\ flux densities.  Similarly, we also measure dust
temperature from the 70\um\ flux densities \citep[as is done in][for
70\um-luminous radio galaxies]{casey09b}, and find that they are
overestimated systematically by 6$\pm$10\,K.  

The severity of these over- and under predicted dust temperatures is
due in part to the sample selection.  Because the sample is selected
at 250\um, it is likely that considering only the 870\um\ or 70\um\
points will produce larger $T_{d}$ error than the 870\um\ or 70\um\
selected samples, simply because of the temperature-weighting and
biasing of these selection wavelengths.  In other words, if a galaxy
is 870\um-bright and is 250\um-faint, then it is far more likely to
have a cooler inherent temperature than a galaxy which is bright at
both wavelengths.  This highlights the difficulty with fitting dust
temperatures to single FIR flux measurements and demonstrates that the
luminosity-temperature distribution of previously studied ULIRG
populations should be revisited when more complete SED information is
gathered from {\it Herschel} and {\sc Scuba-2}
\citep[e.g.][]{magnelli10a}.
%Despite the fact that
%the fitted differences here could be exaggerated by the requirement of
%a 250\um-bright source, the disparity between $L_{FIR}$-$T_{d}$
%derived from only radio/$S_{870}$, radio/$S_{70}$, or direct FIR
%measurements emphasises that there is significant uncertainty in known
%aggregate ULIRG properties to date.

%or full FIR SED information$-$particularly when the FIR/radio
%correlation does a poor job at estimating FIR luminosity$-$emphasises
%the uncertainties in known aggregate ULIRG properties to date.
%Having a unique data set with reliable high S/N FIR measurements, coupled with
%the careful removal of confusion and Eddington bias from the sources' flux densities, 

%Although the measured AGN fraction of high-$z$
%ULIRGs is $\sim$20\%, the high AGN fraction we find is consistent with
%the $>$10$^{13}$\,\lsun\ luminosities of this sample \citep[a similar
%  AGN fraction is observed among the brightest SMGs; e.g.][]{neri03a,
%  greve05a}.

\subsection{HyLIRG Evolution}\label{sec:hylirgevolution}

Figure~\ref{fig:lfirtd} highlights where 250\um\ observations are more
sensitive to hotter dust temperatures than $\sim$870\um\ at $z>1$.
The sparsity of detections in the highlighted region of
Fig.~\ref{fig:lfirtd} indicates that hot-dust HyLIRGs are genuinely
much more rare than their cold-dust analogues at $L_{\rm
FIR}\,>\,10^{13}$\,\lsun\ at high redshift.
%Figure~\ref{fig:lfirtd} highlights the increased sensitivity of 250\um\
%at hotter dust temperatures than 870\um\ observations.  This figure
%highlights a luminosity-temperature area where 250\um\ observations
%are more sensitive than 870\um\ (or similarly, 850\um) at $z\,>$\,1.
%While detecting hotter dust systems is a trade-off with losing the
%sensitivity at lower luminosities where 870\um\ benefits (because of
%limited BLAST depth), 250\um\ observations enable the
%detection of hotter-dust analogues to the SMG population.  Worth
%noting is the sparsity of detections in this area, indicating that
%hot-dust HyLIRGs are genuinely much more rare than their cold-dust
%analogues at $L_{\rm FIR}$\,\simgt\,10$^{13}$\,\lsun\ at high
%redshift.

The dearth of hot-dust ULIRGs (\simgt\,60\,K) from these data is only
significant in the HyLIRG ($>$10$^{13}$\,\lsun) regime for redshifts
above $z\,$=\,1.5 (in other words, it is also significant at fainter
luminosities at lower redshifts but not fainter luminosities at high
redshifts).  Due to the 250\um\ BLAST sensitivity, we have only one
10$^{12}$\,$<$\,$L_{FIR}$\,$<$\,10$^{13}$\,\lsun\ ULIRG in our sample,
and its redshift is $\sim$1.3.  This leaves the possibility that
z$\sim$2 hotter-dust ULIRGs exist, but lie beneath current 250\um\
imaging depth.  \citet{casey09b} showed that at slightly lower
redshifts, z$\sim$1.5, star-formation dominated hot-dust ULIRGs
($T_{d}\sim52$\,K, $L_{\rm FIR}$\simgt\,2$\times$10$^{12}$\,\lsun)
have been observed at 1/5 the volume density of SMGs, but limitations
in $Spitzer$-MIPS 70\um\ depth prevented significant detections at
z\,\simgt\,2.  Also, work by \citet{casey09d} argues that hotter-dust
$>$60\,K HyLIRGs at z$\sim$2 are less prevalent based on CO
observations of submm-faint radio galaxies.  After accounting for
selection bias, the submm-faint ULIRG sample was $\sim$2$\times$ less
luminous in $L_{FIR}$ and $L_{CO}$ than CO-observed, cold-dust SMGs.

If we assume $a\ priori$ that high-$z$ ULIRGs have the same dust
temperature distribution as local ULIRGs (which have $\langle
T_d\rangle =$45$\pm$10\,K above 10$^{13}$\,K), then there is a
$\sim$60\%\ chance that no $>$60\,K sources are detected within a
sample of seven sources (given a Gaussian distribution of dust
temperatures for systems $>$10$^{13}$\,\lsun).  This illustrates how
limiting our sample size is when it comes to drawing conclusions for
the whole 250\um-luminous population.  For example, a sample of
$\sim$30 sources with $T_{\rm dust}\,<\,$60\,K must be detected in
order for that likelihood to drop to \simlt13\%.  More spectroscopic
observations and FIR characterisations of similar samples are needed
from {\it Herschel} and {\sc Scuba-2} for real progress to be made in
high-$z$ ULIRG evolutionary studies, and to probe the differences with
local ULIRG populations.

Despite its significant uncertainty given the small sample size, the
lack of hot-dust systems ($>$60\,K) in the BLAST HyLIRG sample is
consistent with predictions from SPH simulations for infrared luminous
galaxies \citep[see][]{narayanan09a}.  They suggest that the brightest
high-$z$ starbursts ($>$10$^{13}$\,\lsun) are at their most active
phase during the early stages of final merger in-fall, when gas and
dust are diffuse, extended, and cold.  Warmer dust is suggested to
condense either at a later stage merger, when gas and dust has
collapsed and heated, or when it has been heated by a growing AGN.
%Alternatively hotter dust
%may arise in a secularly evolving system or from minor merger
%collisions, when gas and dust are accreted rapidly onto a larger disk
%which remains intact, forming in clumps and ripples where the dust is
%more compact, thus warmer.  
While this sample of galaxies exhibits warm dust (30$<\,T_{\rm
dust}\,<$50\,K), we have not detected any $>$10$^{13}$\,\lsun\
hot-dust ($>$60\,K) systems in this sample and we find a modest AGN
fraction (20\%); therefore, our results loosely support the theory that
the most luminous HyLIRGs are triggered by major merging events.

\subsection{Confusion}\label{sec:confusion}

It is important to once again consider the impact of confusion
limitations and deboosting factors on our conclusions.  We excluded
the uncertainty in the deboosting factor from our results (as
discussed in \S~\ref{s:observations}) because its blind propagation
into all of the bands in our SED fits is not justified.  As is
discussed, the deboosting uncertainty is likely to have far greater
effect on $L_{\rm FIR}$ than on $T_{d}$ or $\beta$ since boosting is
correlated between FIR bands.  Using a naive model where the main
contribution to source confusion is a single additional source within
the beam, whose FIR colours have the same distribution as seen in {\it
  Herschel} populations, we estimate that the derived $L_{\rm FIR}$ is
uncertain by $\sim$0.1\,dex and $T_{\rm dust}$ is uncertain by
$\sim$9\,K.  However, without a proper understanding of the sources
which boost the flux of our high-$z$ BLAST sample, it is very
difficult to determine how the dust temperatures might change,
although it is unlikely that the mean would shift far from the current
mean, 52$\pm$6\,K.  This differential boosting issue should be
investigated carefully with future, large {\it Herschel}
250\um-selected samples.
%The larger
%samples might reveal hot-dust HyLIRGs at z$\sim$2 which we have failed
%to detect in this paper.

The deboosting effect on luminosity is easier to quantify than the
effect on $T_{d}$.  The maximum uncertainty for the deboosting factor
found by \citet{eales09a} is $\sim$50\%, which would propagate to a
factor of $\sim$2 in luminosity.  The mean luminosity of our sample is
$\sim$3$\times$10$^{13}$\,\lsun, a factor of $\sim$5-10 greater than
most high-$z$ ULIRG populations in the literature.  The factor $\sim$2
difference caused by potential deboosting corrections is not
significant in comparison.  These 250\um-bright galaxies are still
``HyLIRGs,'' thus amongst the most luminous, extreme starbursts
measured at high-redshift.

\section{Conclusions}

The redshift identification of these 250\um-bright, z$\sim$2 HyLIRGs
has allowed a characterisation of their near-IR and FIR properties,
leading to the following conclusions:

Near-IR spectroscopy (as probed here by VLT ISAAC) is an efficient way
of identifying redshifts (50-70\%\ success rates) for FIR sources which
have secure photometric redshifts.  The redshift range of our sample
is $z=$1.3-2.6, averaging to $\langle z\rangle =$2.0$\pm$0.4, making
up the z\simgt1 subset of 250\um-bright BLAST galaxies.  We also find
that \Ha\ star formation rates underpredict their FIR SFRs by $\sim$35
times, and we measure metallicities which are in agreement with other
high-$z$ galaxy samples, include those of much lower luminosities but
of similar stellar mass.

Having multiple FIR flux densities available for each object, we fit
FIR blackbody SEDs to each source and constrain $L_{\rm FIR}$,
$T_{d}$, and $\beta$ independent of radio flux density or mid-IR flux
densities.  We find that the FIR/radio correlation holds, but that SMG
composite spectra, when fit to 24\um\ flux densities, do not
successfully describe the FIR properties of this sample.  We measure
FIR luminosities $\sim$3$\times$10$^{13}$\,\lsun, and dust
temperatures averaging $T_{\rm dust}\,=\,$52$\pm$6\,K.  However, we
warn that both of these conclusions are sensitive to the effects of
flux boosting in the FIR, although we estimate that this should not
change $L_{FIR}$ by more than a factor of $\sim$2$\times$ and $T_{\rm
dust}$ beyond its quoted error.

Since 250\um\ selection is more sensitive to the detection of hotter
dust sources than SMG selection (at 850\um), the lack of $>$60\,K
hot-dust galaxies in our sample is potentially an indication that
high-$z$, high-$L$ galaxies are more extended (with diffuse, cool
dust) on a whole than local ULIRGs.  However our small sample size
limits this conclusion to only \simlt40\%\ likelihood.  A lack of
$>$60\,K hot-dust specimens in the BLAST 250\um\ population could be
telling to the galaxies' evolutionary stage; this work highlights the
need for more observations of larger samples of similar and fainter
sources.  FIR mapping at 70-500\um\ from {\it Herschel} and {\sc
  Scuba2} will further select rare and poorly studied high-$z$ ULIRG
populations like the galaxies presented here, and near-IR
spectroscopic observations will enable further redshift identification
of their counterparts, leading to a better characterisation of the
ULIRG phenomenon at high-$z$.

\section*{Acknowledgments}
We thank Rob Ivison and Jim Dunlop for their help in the analysis and
their comments on this manuscript.  We also thank the anonymous
referee for many helpful suggestions which improved the paper.  Based
on observations made with ESO Telescopes under programme numbers
082.A-0890, 083.A-0666, and 084.A-0192.  CMC thanks the Gates
Cambridge Trust, and IRS thanks STFC for support.

\label{lastpage}
\end{document}